\documentclass[aps,prx,showpacs,amsmath,amssymb,superscriptaddress,twocolumn,longbibliography]{revtex4-2}
\usepackage[english]{babel}
\usepackage[utf8]{inputenc}
\usepackage{amsfonts}
\usepackage[T1]{fontenc}
\usepackage[pdftex]{graphicx}
\usepackage{url}
\usepackage[titletoc,title]{appendix}
\usepackage{epstopdf}
\usepackage{amsmath}
\usepackage{amssymb}
\usepackage{dsfont}
\usepackage{mathtools}
\usepackage{xcolor}
\usepackage{tikz}
\usepackage[export]{adjustbox}
\usetikzlibrary{arrows.meta}

\usepackage{hyperref}
\hypersetup{
    bookmarks=false, 
    unicode=false, 
    pdftoolbar=false, 
    pdfmenubar=true, 
    pdffitwindow=false, 
    pdfstartview={FitH}, 
    pdftitle={Optimal steering of matrix product states and quantum many-body scars}, 
    pdfauthor={M. Ljubotina et al.}, 
    pdfsubject={}, 
    pdfcreator={}, 
    pdfproducer={}, 
    pdfkeywords={MPS, TDVP, optimal control, quantum many-body scars}, 
    pdfnewwindow=true, 
    colorlinks=true, 
    linkcolor=black, 
    citecolor=black, 
    filecolor=black, 
    urlcolor=black 
}

\newcommand{\tr}{{\rm Tr}}
\newcommand{\id}{\text{\usefont{U}{bbold}{m}{n}1}}

\newcommand{\mytitle}{Optimal steering of matrix product states and quantum many-body scars}
\begin{document}

    \title{\mytitle} 
    \author{Marko Ljubotina}
    \affiliation{IST Austria, Am Campus 1, 3400 Klosterneuburg, Austria}
    \author{Barbara Roos}
    \affiliation{IST Austria, Am Campus 1, 3400 Klosterneuburg, Austria}
    \author{Dmitry A. Abanin}
    \affiliation{Department of Theoretical Physics, University of Geneva, Quai Ernest-Ansermet 24, 1205 Geneva, Switzerland}
    \author{Maksym Serbyn}
    \affiliation{IST Austria, Am Campus 1, 3400 Klosterneuburg, Austria}
    
    \begin{abstract} 
        Ongoing development of quantum simulators allows for a progressively finer degree of control of quantum many-body systems. 
        This motivates the development of efficient approaches to facilitate the control of such systems and enable the preparation of non-trivial quantum states. 
        Here we formulate an approach to control quantum systems based on matrix product states~(MPS). 
        We compare  counter-diabatic and leakage minimization approaches to the so-called local steering problem, that consists in finding the best value of the control parameters for generating a unitary evolution of the specific MPS state in a given direction. 
        In order to benchmark the different approaches, we apply them to the generalization of the PXP model known to exhibit coherent quantum dynamics due to quantum many-body scars. 
        We find that the leakage-based approach generally outperforms the counter-diabatic framework and use it to construct a Floquet model with quantum scars. 
        We perform the first steps towards global trajectory optimization and demonstrate entanglement steering capabilities in the generalized PXP model. 
	    Finally we apply our leakage minimization approach to construct quantum scars in the periodically driven non-integrable Ising model. 
    \end{abstract}
    \maketitle

    \section{Introduction}
    \label{sec:introduction}

    Robust manipulation and preparation of quantum states in isolated interacting quantum systems remains an outstanding problem. 
    Numerous platforms such as superconducting qubits~\cite{Houck:2012vz,Devoret13}, nitrogen-vacancy centers~\cite{DOHERTY20131}, cold atomic systems~\cite{Bloch2008,lewenstein2012ultracold}, trapped ions~\cite{Blatt2012}, and Rydberg atom arrays~\cite{Labuhn2016,Bernien2017,Browaeys2020,Dolev20} provide access to non-equilibrium quantum dynamics governed by a known (effective) Hamiltonian with a handful of experimentally controllable parameters. 
    Despite a number of successful examples~\cite{RevModPhys.76.1037,Frank:2016uk,Zhou:2017wc}, our understanding of optimal ways for preparing potentially useful quantum states in these platforms remains limited. 
    
    Indeed, even numerical simulation of non-equilibrium dynamics of quantum many-body systems represents a challenge due to a Hilbert space dimension that grows exponentially with the number of degrees of freedom. 
    The challenge related to the exponential resources needed to store the many-body wave function can be mitigated by tensor-network based approaches. 
    For one dimensional systems, matrix product states (MPS) provide an efficient representation of quantum states that have area-law scaling of entanglement entropy~\cite{PerezGarcia2007,Schollwock}. 
    The MPS representation was employed in numerical approaches to optimize state preparation in quantum systems, see Refs.~\cite{Doria11} along with review~\cite{CRABreview} and tutorial~\cite{Boscain21}. 
    Typically, in such numerical setups one optimizes the fidelity of the state preparation over the trajectory in the multi-dimensional space of control parameters using gradient-free~\cite{CRAB} or gradient-based routines~\cite{PhysRevA.95.012317,PhysRevA.84.022305,Jesper21,Jesper21-2}. 
    In addition, machine-learning based approaches to this problem were also considered~\cite{Bukov18,Bukov22}. 

    An alternative approach to state manipulation is provided by the adiabatic state preparation procedure and its modification. 
    This procedure relies on the adiabatic theorem~\cite{adiabatic} that guarantees that a system remains in its ground state under a sufficiently slow change of Hamiltonian parameters, provided the system remains gapped. 
    Several approaches aimed at speeding up the adiabatic procedure without loosing performance, known as ``shortcuts to adiabaticity'' were proposed~\cite{Demirplak:2003wf,Berry_2009,Campo12,Campo13,Opat14,Saber14,PhysRevA.96.013431,PhysRevX.4.021013,sels2017,PhysRevA.98.043436}. 
    These approaches either allow to create and then coherently remove excitations during state preparation, or use additional control fields to implement the so-called counter-diabatic (CD) terms~\cite{Demirplak:2003wf,sels2017}, which ensure the absence of excitations at each point in time. 
    We note that these adiabatic approaches and their extensions typically deal with finding the optimal control \emph{along a fixed trajectory} in the parameter space. 
    
    Both numerical optimization and adiabaticity-based approaches discussed above come with their own challenges. 
    On the one hand, the physics behind the optimal solution provided by the purely numerical approaches often remains unclear. 
    On the other hand, approaches that build upon modification of adiabatic state preparation are typically variational and often lack an efficient implementation. Furthermore, it is difficult to quantify their performance. 
    
    In this work we provide an alternative approach to the problem of quantum state control that resolves some of the above issues. 
    Similarly to the numerical approaches~\cite{Doria11}, we use the manifold of MPS states for efficient representation of wave functions and the desired trajectory of quantum state evolution. 
    We provide an efficient solution to the problem of ``optimal steering of states within an MPS manifold'': Given a fixed MPS state, a desired direction of unitary evolution in its tangent space and an available set of control parameters, we find the optimal values of control parameters for steering the many-body wave function along the desired direction. 
    This solution is obtained by minimizing a quantum leakage, that allows to quantify the efficiency of the control locally at each point of the trajectory. 
    Moreover, we discuss the relation between our approach and counter-diabatic driving, demonstrating that the latter can be related to the leakage minimization procedure with a modified cost function. 
    
    In order to illustrate and compare our approach to counter-diabatic driving~\cite{sels2017}, we consider a generalization of the so-called PXP model~\cite{LesanovskyMPS,Bernien2017,Turner2017} that provides an approximate description for dynamics of Rydberg atom arrays~\cite{Bernien2017}. 
    The PXP model is known to exhibit many-body revivals that were attributed to the existence of non-thermal ``quantum scarred'' eigenstates~\cite{Turner2017}. 
    In addition, the dynamics of this model for some initial states can be efficiently described by a low bond dimension MPS manifold~\cite{wenwei18TDVPscar,Michailidis2020}, thus providing a natural model to test our control protocol. 
    
    We find that our leakage-based approach to optimal steering of the generalized PXP model typically outperforms counter-diabatic driving. 
    We introduce the notion of leakage landscape and optimal direction for unitary evolution. 
    These concepts are used to demonstrate entanglement control in the generalized PXP model and may be viewed as steps towards the complete optimization of the trajectory in the parameter space using our method.  
    
    Effectively, our approach maps a quantum control problem onto a classical one, thus enabling a broad range applications. 
    In the context of the generalized PXP model, we construct continuously-driven Floquet-type models with quantum many-body scars. 
    Although scars in continuously driven Rydberg arrays were observed experimentally~\cite{Dolev20}, their theoretical explanation uses Floquet models with pulsed driving~\cite{Nishad21}. 
    Our work introduces a framework for the systematic construction of Floquet models with continuous driving. 
    This opens the door to realizations of quantum scars using quantum simulation platforms beyond Rydberg arrays~\cite{Kao,Scherg:2021tl,WW21,Moudgalya21}, as we demonstrate by constructing a Floquet scar in the Ising model with transverse and longitudinal fields. 
    
    The remainder of the paper is structured as follows. 
    In Sec.~\ref{sec:model} we formulate the local steering problem and discuss a generalization of the PXP model that will be used for benchmarking the different approaches. 
    Next, Sec.~\ref{sec:control} introduces the CD-based and leakage minimization approaches for the local steering problem. 
    In this section we show that our leakage based approach outperforms the CD approaches when applied to the generalized PXP model. 
    Sec.~\ref{sec:landscape}  focuses on the leakage based approach and studies the efficiency of the steering of the quantum system depending on the direction and position within the MPS manifold. 
    The leakage landscape obtained in this section provides an intuition for the trajectory optimization, allowing us to control the entanglement during coherent quantum evolution. 
    In Sec.~\ref{sec:ising} we apply our approach to improve periodic revivals in the driven Ising model with transverse and longitudinal fields, effectively constructing a Floquet scar which we further stabilize by including additional terms in the Hamiltonian. 
    Finally, in Sec.~\ref{sec:discussion} we summarize our findings and discuss the most promising directions for future research. 
    The paper is concluded by Appendices~\ref{app:cd}-\ref{app:ising} that present details of analytical calculations and additional numerical data. 
   
    \section{Control problem and specific model}
    
       \begin{figure}[t]
        \centering
        \includegraphics[width=0.99\columnwidth]{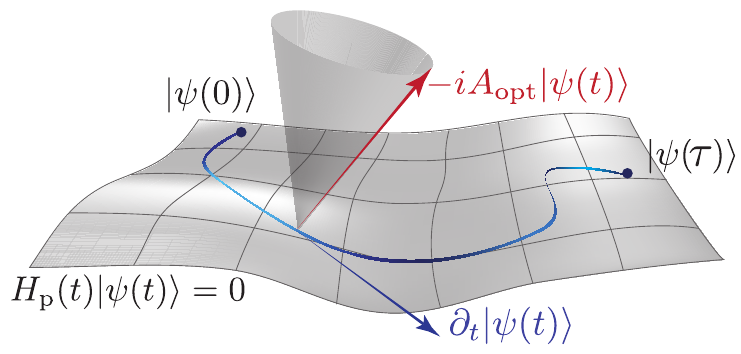}
        \caption{
            Schematic representation of the desired trajectory $|\psi(t)\rangle$ connecting two points $|\psi(0)\rangle$ and $|\psi(\tau)\rangle$ within the MPS manifold $\Psi$. 
            For the MPS manifold one can define a family of parent Hamiltonians $H_\text{p}(t)$ that have a state  $|\psi(t)\rangle$ as their zero energy ground state. 
            The unitary dynamics is generated by the operator $A$ that depends on the control parameters.
            We seek to identify the optimal value of the control parameters at each point on the trajectory, such that the exact unitary dynamics generated by $A_\text{opt}$ is closest to the desired direction of evolution $\partial_t |\psi(t)\rangle$. 
        }
        \label{fig:schematic}
    \end{figure}
    
    \label{sec:model}
    In this Section we first provide a general formulation of the control problem that is addressed within our work. 
    Afterwards we provide an example of such a control problem using a generalization of the PXP model. 
    
    \subsection{Optimal evolution of the state along a trajectory}
    
    In this work we address the problem of optimally steering a quantum many-body system along a certain trajectory. 
    We start by specifying a manifold of many-body wave functions $\Psi$ and a desired wave function trajectory $|\psi(t)\rangle$, see Fig.~\ref{fig:schematic}. 
    We assume that the many-body wave function is subject to unitary evolution generated by the operator $A$, 
    \begin{equation}
        \label{Eq:dynamics}
        \partial_t | \psi_{\rm Q}(t)\rangle = -i A(t) | \psi_{\rm Q}(t)\rangle,
        \  
        A(t)=\sum_\eta c_\eta(t) A_\eta,
    \end{equation}
    where $A$ depends on some control parameters $c_{1}$, $c_2$,\ldots and may be viewed as a time-dependent Hamiltonian of the system. 
    The wave function with index Q, $|\psi_{\rm Q}(t)\rangle$ in what follows will be used to distinguish the exact unitary evolution from the desired MPS trajectory $|\psi(t)\rangle$. 
    Given a trajectory $|\psi(t)\rangle$ and a set of control parameters $c_\eta$, we seek to find the values of these control parameters such that the full quantum unitary evolution $|\psi_{\rm Q}(t)\rangle$ generated by operator $A$ is as close as possible to the desired evolution. 
    
    The formulation of the optimal control problem provided above bears similarity to the variational principle for quantum dynamics~\cite{Dirac1930}.
    The variational approach to quantum dynamics solves the problem of finding the optimal direction in the variational manifold that provides the best approximation to a unitary dynamics generated by a fixed operator $A$.
    In the case when the manifold of quantum states is represented as an MPS, the solution to this problem is provided by the so-called time dependent variational principle for MPS manifolds (abbreviated as TDVP in what follows)~\cite{Haegeman}. 
    Here we consider an inverse problem: we fix the desired direction of unitary evolution to a tangent vector  $\partial_t |\psi(t)\rangle$~(see Fig.~\ref{fig:schematic}) and search for the best generator that approximates such unitary dynamics.
    Out of the linear space of all possible directions that can be generated by varying values of parameters $c_\eta$, schematically shown as a shaded cone in Fig.~\ref{fig:schematic}, we choose the values of $c_\eta$ such that the corresponding generator of dynamics, $A_\text{opt}$, steers the quantum system in the direction that is closest to the desired tangent vector.
    
    This optimal control problem can be solved in several ways discussed in Section~\ref{sec:control} below. 
    However, before reviewing the existing recipes and presenting our approach, we specify a particular example of the operator $A$ and the variational manifold. 
 
    \subsection{Generalized PXP model and variational manifold}
    
    We illustrate different approaches to the optimal control problem using the generalization of the PXP model, which provides an approximate description of dynamics in Rydberg atom quantum simulator~\cite{LesanovskyMPS,Bernien2017,Turner2017}. 
    This model is defined on a one-dimensional chain of effective spin-1/2 degrees of freedom. 
    The PXP Hamiltonian reads 
    \begin{equation}
        \label{Eq:PXP}
        H=\sum_{i=1}^{l}\tilde \sigma^x_{i},
        \qquad \tilde \sigma^x_{i} = P_{i-1}\sigma^x_{i} P_{i+1},
    \end{equation} 
    where $\sigma_i^\alpha$ is the corresponding Pauli matrix acting on the spin on site $i$, whereas $\tilde \sigma_i^\alpha$ is a shorthand notation for this matrix dressed by projector operators on two adjacent sites, $P_{i\pm 1}=|\hspace{-1mm}\downarrow\rangle_{i\pm1}\langle\downarrow\hspace{-1mm}|_{i\pm1}$. 
    This projector enforces the constraint preventing two neighboring sites to be simultaneously in the $\uparrow$-state, which naturally arises due to the Rydberg blockade~\cite{Lesanovsky2011}. 
    The length of the system $l$ is assumed to be infinite throughout this work, except for benchmarking against the exact quantum dynamics, when periodic boundary conditions are used and the value of $l$ is explicitly indicated.

    The PXP model is characterized by an unusual dynamics for certain initial states, notably the N\'eel initial state, $|Z_2\rangle = |\hspace{-1mm}\uparrow\downarrow\uparrow\dots\rangle$. 
    Although the initial $|Z_2\rangle$ state corresponds to a highly excited state of the PXP Hamiltonian~\cite{Turner2017}, the dynamics under the PXP Hamiltonian leads to coherent oscillations between the $|Z_2\rangle$ state and its translated version $|Z'_2\rangle=|\hspace{-1mm}\downarrow\uparrow\downarrow\dots\rangle$.
    This dynamics was attributed to the existence of non-thermal eigenstates embedded throughout the otherwise thermal spectrum, known as quantum many-body scars~\cite{Turner2017}.
    
    Motivated by the scar dynamics that requires a two-site unit cell to describe oscillations between the $|Z_2\rangle$ and $|Z_2'\rangle$ states, we consider the following generator of dynamics,
    \begin{equation}
        A=\sum_{i=0}^{l/2-1}\left(c_1(t)\tilde\sigma^x_{2i+1} + c_2(t) \tilde \sigma^x_{2i+2}\right),
        \label{eq:pxp}
    \end{equation}
    where we use the definition of Pauli matrices dressed by projectors from Eq.~\eqref{Eq:PXP}.
    This generator can be viewed as a generalization of the PXP Hamiltonian that now depends on two control parameters $c_{1,2}$, which are proportional to the Rabi frequency on the even and odd sites of the lattice. 
    While such control of the Rabi frequency is not available in the existing experimental setups~\cite{Dolev20}, these capabilities may be feasible in the future.  
    
    The second component required to fully specify the control problem is a variational manifold of states $\Psi$, here chosen to be an MPS manifold,
    \begin{equation}\label{Eq:MPS-general}
        \Psi = \{|\psi(x_1,\dots)\rangle=\tr\left[\prod_{i=1}^{l} M_i(x_1,\dots)\right];~x_j\in\mathbb{R}\},
    \end{equation}
    where $M_i$ are the matrices defining the MPS state and $x_j$ are some real parameters.
    Consistent with the choice of our generator of dynamics, $A$, we choose an MPS that has a two-site unit cell. 
    A natural choice is provided by the following bond dimension $\chi=2$ ansatz~\cite{Bernien2017,wenwei18TDVPscar,Michailidis2020}
    \begin{equation}
        M_i=\begin{pmatrix}\cos\theta_i|\hspace{-0.9mm}\downarrow\rangle_i & i\sin\theta_i
        |\hspace{-0.9mm}\uparrow\rangle_i\\|\hspace{-0.9mm}\downarrow\rangle_i & 0\end{pmatrix},
        \label{eq:mps}
    \end{equation}
    where $\theta_1$ and $\theta_2$ are the two free real parameters corresponding to even and odd sites respectively.
    Defining a trajectory within the MPS manifold is then equivalent to defining $\theta_i\equiv\theta_i(t)$. 
    
    This ansatz suffices for an accurate description of the dynamics generated by the PXP model starting from the $|Z_2\rangle$ state. 
    In particular, the TDVP projection~\cite{Haegeman} of the PXP model dynamics onto this manifold yields a nearly circular periodic trajectory that connects the $|Z_2\rangle$ and $|Z_2'\rangle$ states~\cite{Bernien2017,wenwei18TDVPscar} that correspond to values of angles $(\theta_1,\theta_2) = (\pi/2+k\pi,\vartheta)$ and $(\vartheta,\pi/2+k\pi)$, respectively.
    Here $\vartheta$ is a free parameter that can take any real value except for $\vartheta=\pi/2+m\pi$ which represents the singular state $(|Z_2\rangle+|Z'_2\rangle)/{\sqrt{2}}$ disconnected from the rest of the manifold. 
    
    Finally, the above choice of MPS will allow us to define a relatively simple parent Hamiltonian~\cite{LesanovskyMPS}, which we will use for the CD approaches below. 
    This parent Hamiltonian can be conveniently written via a local Hamiltonian density operator that depends on two coupling constants, $h^\text{p}(a,b)$,
    \begin{equation}
        h^{\rm p}_i(a,b)=a \tilde\sigma^y_{i}+b \tilde P_{i} + \frac{a^2}{b} \tilde n_{i},
        \label{eq:parent-density}
    \end{equation}
    where the operator $n_i=\id-P_i=|\hspace{-1.2mm}\uparrow\rangle_i\langle\uparrow\hspace{-1.2mm}|_i$ is the projector complementary to $P_i$, and tilde above operators denotes dressing them with projectors as in Eq.~\eqref{Eq:PXP}: for example, $ \tilde P_{i}  = P_{i-1} P_{i} P_{i+1} $. 
    Written in this notation, the parent Hamiltonian for the MPS state \eqref{Eq:MPS-general}-\eqref{eq:mps} takes the following form,
    \begin{equation}
        H_{\rm p}=\sum_{i=0}^{l/2-1}\left[h^{\rm p}_{2i+1} (a_1,b_1) +h^{\rm p}_{2i} (a_2,b_2) \right].
        \label{eq:parent-general}
    \end{equation}
    The parameters $a_{1,2}$ and $b_{1,2}$ that specify the Hamiltonian density on odd and even lattice sites depend on the values of angles $\theta_{1,2}$ as $b_{1,2}=\pm a_{1,2}\frac{\tan\theta_{1,2}}{\cos\theta_{2,1}}$.
    This Hamiltonian is frustration free, which implies that the ground state of $H_{\rm p}$ is simultaneously also the ground state of all $h^{\rm p}_i$. 
    For our MPS to be the ground state of this Hamiltonian, we must have $b_i>0$, whereas for $b_i<0$ the MPS state becomes the highest excited state. 
    In practice, we can fix $a_{1,2}=1$ for simplicity, since the absolute value of $a$ is not relevant for the structure of the ground state, merely representing an overall scale in the system. 
    
    \section{Counter-diabatic and leakage approaches to quantum control}
    \label{sec:control}
    
    In this section we review the variational counter-diabatic approach~\cite{sels2017} and apply this approach to the problem of steering a many-body quantum system along a given trajectory. 
    Afterwards, we formulate an alternative approach that, similar to TDVP, relies on the concept of leakage minimization. 
    Finally, in Sec.~\ref{sec:compare} we compare different approaches using the generalized PXP model introduced in the previous section. 
    
    \subsection{Counter-diabatic driving}
    \label{sec:CD}

    \begin{figure*}[t]
        \centering
        \includegraphics[width=\textwidth]{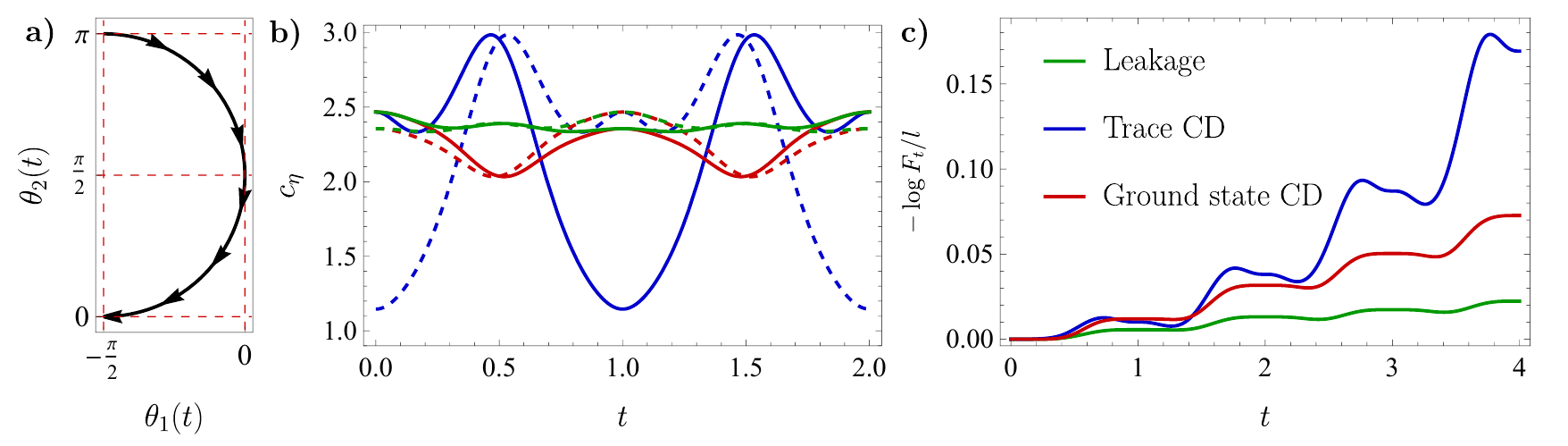}
        \caption{
            We compare different approaches to the control problem shown in Figure 1.
            Panel (a) shows the desired trajectory in the plane of angles $\theta_{1},\theta_2$ that parametrize the MPS state, that closely resembles the scarred trajectory~\cite{Bernien2017,wenwei18TDVPscar}.
            Panel (b) shows the optimal value of driving parameters obtained from trace CD (blue), ground state CD (red) and leakage-based (green) optimization. Full (dashed) lines correspond to $c_1$ ($c_2$) respectively.
            The leakage approach yields a nearly constant value of driving parameters along the full trajectory.
            Finally, panel (c) shows the time dependence of the logarithm of fidelity ($F_t=|\langle\psi(t)|\psi_{\rm Q}(t)\rangle|^2$) per site computed using the iTEBD approach~\cite{Vidal07}, illustrating that the leakage approach outperforms CD-based approaches. 
            All data shown were computed using a bond dimension of $\chi=64$ and a time step $\delta t=10^{-5}$. 
        }
        \label{fig:comp}
    \end{figure*}
    
    The CD approach aims to minimize transitions from the ground state to the excited states of a time-dependent Hamiltonian. 
    The idea is to add additional terms $A(t)$, called the adiabatic gauge potential, to the time dependent Hamiltonian, $H_\text{p}(t)$, that depend on variational parameters as defined in Eq.~\eqref{Eq:dynamics}. 
    The CD approach of Ref.~\cite{sels2017} provides a specific recipe for determining the optimal values of parameters $c_\eta$ at each point in time. 
    
    We extend the CD approach to the control problem over MPS manifold using two observations. 
    First, for any MPS state there exists a local parent Hamiltonian~\cite{Fannes1992,PerezGarcia2007}.
    Hence, the trajectory in Fig.~\ref{fig:schematic}, $|\psi(t)\rangle$ specifies a time-dependent parent Hamiltonian $H_\text{p}(t)$. 
    Second, the condition that the parent Hamiltonian has an MPS state as its zero energy ground state, $H_\text{p}(t)|\psi(t)\rangle = 0$, that can be additionally imposed, allows us to eliminate the parent Hamiltonian from the unitary dynamics~\footnote{We note that such elimination of $H_\text{p}(t)$ is exact in the limit when the state $|\psi_{\rm Q}(t)\rangle=|\psi(t)\rangle$. Furthermore, there is ambiguity in $H_{\rm p}(t)$, not least since we can choose the norm arbitrarily $|H_{\rm p}(t)|=\epsilon$ without affecting the driving parameters $c_\eta$.}, leaving only $A$ in the time-dependent Schr\"odinger equation, $\partial_t |\psi_{\rm Q}(t)\rangle=-i A(t)|\psi_{\rm Q}(t)\rangle$, that thus becomes equivalent to Eq.~\eqref{Eq:dynamics}. 
    
    The above observations allow us to apply the CD approach to our control problem. 
    Using the parent Hamiltonian for each point in the MPS manifold of states, we define an operator
        \begin{equation}
            \label{eq:CD-G}
            G(t)=\partial_tH_\text{p}(t)+i[A(t),H_\text{p}(t)],
        \end{equation}
    that can be viewed as a ``force'' that causes the transitions between the instantaneous eigenstates of $H_\text{p}(t)$ in the process of the time evolution. 
    The essence of the CD approach is to find the best values of parameters $c_\eta(t)$ in the operator $A$, such that transitions described by the operator $G(t)$ are minimized.
    
    If the ground state of the Hamiltonian $H_\text{p}(t)$ is not known, it is natural to resort to the minimization of the operator norm of $G(t)$, an approach that we call ``trace-CD''~\cite{sels2017}. 
    In the trace-CD approach the trace of the square of $G(t)$ gives a cost function, 
    \begin{equation}
        \label{eq:CD-inf}
        S(t)=\tr\left[G(t)^2\right],
    \end{equation}
    that is minimized at each point in time. 
    Minimizing the trace of the operator $G(t)$ treats all the states in the many-body spectrum on equal footing, corresponding to an infinite-temperature density matrix of the system. 
    While this approach offers analytic tractability for local controls, the infinite temperature optimization does not yield the best results for the fidelity of remaining in the target state.

    To apply the trace-CD approach to the PXP model, we calculate operator $G(t)$ explicitly, using Eq.~\eqref{eq:pxp} and the expressions of the parent Hamiltonian \eqref{eq:parent-density}-\eqref{eq:parent-general}. 
    Writing this operator as a sum of local terms, 
    \begin{equation}
        G(t)=\sum_{i=1}^{l/2}(G^{(1)}_i(t)+G^{(2)}_i(t)),
    \end{equation}
    we get
    \begin{multline}
        G^{(1)}_i(t)=\left(\frac{2a_1a'_2}{b_1}-2a_1c_1-\frac{a_1^2b'_1}{b_1^2}\right){\tilde{\id}_{2i}}
        \\
        +\left(4a_1c_1-\frac{2a_1a'_1}{b_1}+b'_1+\frac{a_1^2b'_1}{b_1^2}\right){\tilde{P}_{2i}}
        \\+\left(\frac{a_1^2+b_1^2}{b_1}c_1+a_1'\right){\tilde{\sigma}_{2i}^y}
        -b_1c_2\left({\tilde \sigma}_{2i-1}^y P_{2i+1}+{P_{2i-1}\tilde \sigma_{2i+1}^y}\right)
        \\
        -(a_1c_2+a_2c_1)\left(P_{2i-2}{\sigma_{2i-1}^+\sigma_{2i}^-}P_{2i+1}+\text{H.c.}\right).
    \end{multline}
    $G^{(2)}_i(t)$ is obtained by changing $1\leftrightarrow2$ for all $a,b,c$ and shifting all positions by $1$. 
    Here $\tilde{O}$ represents an operator $O$ dressed by projectors $P$ on the left and right. 
    The cost function $S(t)$ can be calculated using the values of the traces of Pauli matrices calculated over the constrained Hilbert space (see Appendix~\ref{app:trace-calc} for details). 
    This cost function is a quadratic polynomial in $c_{1,2}(t)$ whose minimization is straightforward. 
    
    We consider a circular trajectory from Fig.~\ref{fig:comp}a, defined as 
    \begin{equation}\label{Eq:circle}
        \theta_1(t)=\frac{\pi}{2}(\sin\pi T(t)-1), 
        \qquad \theta_2(t)=\frac{\pi}{2}(\cos\pi T(t)+1),
    \end{equation}
    which is close to the scarred trajectory of the PXP model~\cite{Bernien2017,wenwei18TDVPscar}. 
    Here $T(t)$ is a function with $T(0)=0$ and $T(\tau)=1$ that specifies both the time taken by the whole procedure and the instantaneous velocity at which we travel along the trajectory at each point in time~\footnote{Note that since $c_\eta\propto T'(t)$, our results for trace-CD and all other approaches are only trivially affected by such rescaling of time. As a result, we fix $T(t)=t/\tau$ with $\tau=1$ throughout this paper.}. 
    The optimal values of the parameters $c_{1,2}$ resulting from the minimization of $S(t)$ along the circle trajectory are illustrated in Fig.~\ref{fig:comp}b (blue color). 
    
    Alternatively, the knowledge of the desired state $|\psi(t)\rangle$ in the MPS form allows us to define the cost function using the expectation value over the state $|\psi(t)\rangle$, instead of a trace, 
    \begin{equation}
        \label{eq:CD-zero}
        S(t)=\langle\psi(t)|G(t)^2|\psi(t)\rangle-\langle\psi(t)|G(t)|\psi(t)\rangle^2,
    \end{equation}
    where in Appendix~\ref{app:GS} we show that the contribution from the second term vanishes when the energy of $|\psi(t)\rangle$ is chosen to be zero at all times, $H_\text{p}(t)|\psi(t)\rangle=0$.
    We will refer to this method as the ``ground state CD'' approach, which can be viewed as performing the same calculation as before using a density matrix corresponding to the pure state $|\psi(t)\rangle$. 
    Importantly, one can view this approach as only minimizing transitions from the ground state as opposed to all states in the trace-CD approach~\cite{sels2017}, naturally leading to better performance. 
    
    In general, the ground state CD approach requires more involved calculations compared to the trace-CD approach. 
    Naively Eq.~\eqref{eq:CD-zero} includes the expectation values of products of up to four different local operators over a given MPS state.
    While such four point correlation functions can be computed efficiently, below we formulate specific conditions that greatly simplify the evaluation of Eq.~\eqref{eq:CD-zero}. 
    When both the parent Hamiltonian and gauge potential can be represented as a sum of local operators [i.e.\ local energy density like in Eq.~\eqref{eq:parent-density}], it can be shown that it is sufficient to evaluate simple expectation value of these operators (one-point correlation functions). 
    While higher correlation functions still enter the expression for the cost function, Eq.~\eqref{eq:CD-zero}, they do not contribute to the optimization with respect to $c_\eta(t)$. 
    
    In Appendix~\ref{app:GS} we present the details for the ground state CD approach applied to the generalized PXP model. 
    In our example, the analytic expressions for $c_{1,2}(t)$ found within this approach are cumbersome and are listed in~\cite{Note3}. 
    The resulting behavior of $c_{1,2}(t)$ for the circular trajectory in Fig.~\ref{fig:comp}b (red lines). 
    differs significantly from case of trace CD optimization.
    Remarkably, the coefficients  $c_{1,2}(t)$ obtained from the ground state CD approach reveal much weaker dependence on the position within the trajectory, changing by less than $20\%$ along the full trajectory. 
    
    \subsection{Leakage minimization}
    \label{sec:leak}
    
    After reviewing solutions provided by the CD approaches, we formulate an alternative framework inspired by TDVP. 
    The TDVP is obtained by minimization of the disagreement between the exact quantum evolution and its projection onto the variational manifold. 
    This disagreement is quantified by the so-called \emph{leakage} --- the norm of the vector given by the difference between the exact quantum evolution $-iA(t)|\psi(t)\rangle$ and its projection onto the variational manifold,  $\partial_t|\psi(t)\rangle$. 
    This leads to the following definition of the leakage,
    \begin{equation}
        \label{eq:leakage}
        \delta^2(t)=\frac{1}{l} {\big|\partial_t|\psi(t)\rangle+iA(t)|\psi(t)\rangle\big|^2},
    \end{equation}
    where we included an additional factor of system size, $l$, to make this quantity well behaved in the thermodynamic limit $l\to\infty.$ 
    The standard TDVP equations  are obtained from minimizing $\delta^2(t)$ over $\partial_t|\psi(t)\rangle$~\cite{Haegeman}. 
    
    Our key insight is that minimizing the leakage over parameters in $A(t)$ provides an efficient solution to the control problem specified in Section~\ref{sec:model}. 
    When $A(t)$ is given by Eq.~(\ref{Eq:dynamics}), $\delta^2(t)$ becomes a quadratic polynomial in~$c_\eta$, 
    \begin{eqnarray}
        \delta^2(t) &=& c_\eta D_{\eta \rho}c_\rho+e_\eta c_\eta+{\rm const.},\\
        \label{eq:D}
        D_{\eta \rho}&=&\langle\psi(t)|\{A_\eta,A_\rho\}|\psi(t)\rangle_c,\\
        \label{eq:e}
        e_{\eta}&=&2{\rm Im}\langle \psi(t)| A_\eta{\partial_t}|\psi(t)\rangle_c,
    \end{eqnarray}
    where summation over repeated indices is implied and  connected correlation functions are defined in a standard way, $\langle \psi(t)| O_1O_2|\psi(t)\rangle_c = \langle \psi(t)| O_1O_2|\psi(t)\rangle- \langle \psi(t)| O_1|\psi(t)\rangle \langle \psi(t)|O_2|\psi(t)\rangle$.
    Minimization of quadratic polynomial can be easily performed analytically, yielding an explicit answer for the optimal value of driving parameters,
    \begin{equation}\label{Eq:c-general-leakage}
        c_\eta=-D^{-1}_{\eta \rho}e_\rho,
    \end{equation}
    where $D^{-1}$ is the pseudo-inverse of the matrix $D$, defined such that the kernels of $D$ and $D^{-1}$ are the same. 
   
    We note that TDVP minimization and the present minimization of leakage can be related to each other. 
    We delegate the detailed comparison between these to Appendix~\ref{app:leak-vs-tdvp}, and mention only their qualitative difference. 
    Qualitatively, in the TDVP approach one finds the optimal projection of the unitary evolution onto the tangent space, thus TDVP leakage penalizes only components of $-iA |\psi(t)\rangle$ that are orthogonal to the tangent space. 
    Instead, our approach minimizes the leakage over all possible generators of the unitary dynamics, thus effectively penalizing any components $-iA|\psi(t)\rangle$ (including those that belong to the tangent space of the MPS manifold) that are orthogonal to the desired direction of evolution, $\partial_t|\psi(t)\rangle$.

    Finally, before applying this framework to the PXP model, we discuss its relation to the CD approach from Sec.~\ref{sec:CD}. 
    The trace-CD approach deals with the entire Hilbert space and thus cannot be immediately related to the present procedure. 
    However, the ground state CD approach can be viewed as an optimization of the weighted leakage. 
    Specifically, in Appendix~\ref{app:leak-vs-GS} we show that ground state CD can be viewed as a minimization of the following cost function,
    $S(t)=|H_\text{p}(t)\left[\partial_t+iA(t)\right]|\psi(t)\rangle|^2$.
    This cost function represents the leakage vector that is multiplied by the instantaneous frustration-free parent Hamiltonian. 
    The application of parent Hamiltonian, penalizes leakage to different excited states with a different weight: the closer the state is to the ground state, the smaller the penalty. 
    While weighting of the leakage with the parent Hamiltonian could be physically motivated, we believe that leakage minimization approach is more natural since it does not require the notion of a parent Hamiltonian, which need not be unique.

    Applying the leakage approach to the optimal control problem of the generalized PXP model requires explicit evaluation of one- and two-point correlation functions that is described in Appendix~\ref{app:leak-calc}.
    This calculation leads to the following explicit form for the values of couplings $c_{1,2}(t)$, 
    \begin{equation}   \label{eq:leakC}
        c_{1,2} =\frac{n_{1,2} \theta'_{1,2}(t) - m_{1,2} \theta'_{2,1}(t)}{f_{1,2}},
    \end{equation}
    where the constants $f_{1,2}$, $n_{1,2}$ and $m_{1,2}$ are complex trigonometric expressions related to the expectation values of different operators over the MPS state~\eqref{eq:mps}. 
    Specifically, they read
    $n_{1} =2(9+6\cos2\theta_{1}+\cos4\theta_{1})\cos\theta_{2}+8\cos2\theta_{1}\cos3\theta_{2}\sin^2\theta_{1}$, $
    m_{1} = (2\sin2\theta_{1}+7\sin4\theta_{1})\sin\theta_{2}+8\cos\theta_{1}\sin^3\theta_{1}\sin3\theta_{2}
    $, and, finally
    $f_{1} = -[29+3\cos4\theta_{1}+32\cos2\theta_{1}\cos2\theta_{2}+6\cos4\theta_{2}\sin^2 2\theta_{1}\Big]/2$ with constants $n_2$, $m_2$ and $f_2$ obtained by swapping indices $1\leftrightarrow 2 $.
    
    An example of these solutions for the circular trajectory used before is shown in Fig.~\ref{fig:comp}(b) by the green line.
    Note that the resulting driving parameters are nearly flat in the vicinity of the scar trajectory. 
    Indeed, one could also use this approach to look for scars in other Hamiltonian models, since nearly constant-in-time ratios of driving parameters might signify the presence of a scar~\footnote{Nearly constant-in-time ratios of driving parameters essentially correspond to our model being a small time-dependent perturbation of a time-independent Hamiltonian model. Nevertheless, this is neither a necessary nor sufficient condition for the nearby Hamiltonian model to have a scar corresponding to the chosen trajectory.}. 
    This implies that the PXP model, where values $c_{1,2}(t)$ are fixed to be constants, may be potentially viewed as a small deformation of a nearby Floquet model with an exact eigenstate corresponding to the scar trajectory. 
    
    Finally, we define a natural geometric quantity which can be viewed as a properly normalized leakage,
    \begin{equation}
        \label{eq:leakAngle}
        \Delta^2=\frac{1}{l} \frac{\big|\partial_t|\psi(t)\rangle+iA(t)|\psi(t)\rangle\big|^2}{\big|\partial_t|\psi(t)\rangle\big|^2},
    \end{equation}
    and corresponds to $\sin^2\alpha$, where $\alpha$ is the angle between the optimal direction $-iA(t)|\psi(t)\rangle$ in the manifold spanned by $-iA(\{c_j\})|\psi(t)\rangle$ and the desired trajectory $\partial_t|\psi(t)\rangle$, see Fig.~\ref{fig:schematic}. 
    This quantity assumes values between 0 and 1, with zero corresponding to the absence of any leakage, and one corresponding to the case when the ``optimal'' direction is nevertheless fully orthogonal to the desired direction. 
    This normalized leakage will be used when determining an optimal trajectory between two points in the MPS manifold in Sec.~\ref{sec:landscape}.
    
    \subsection{Comparing different approaches}
    \label{sec:compare}
    
    Thus far we have formulated two CD-based approaches and the leakage minimization approach to the control problem. 
    All these approaches are variational, and in this section we benchmark their efficiency.
    To this end, we simulate the exact quantum dynamics, $|\psi_{\rm Q}(t)\rangle = {\cal T}e^{-i\int_0^td t'\,A(t')}|\psi(0)\rangle$ induced by the time-dependent $A(t)$ found by each of the approaches above. 
    In order to compare the results we look at the fidelity density
    \begin{equation}
        f(t)=-\frac{1}{l}\log\left|\langle\psi(t)|\psi_{\rm Q}(t)\rangle\right|^2,
        \label{eq:fidelity}
    \end{equation}
    which provides a system-size independent measure of how close to the desired state we are at any point in time. 
    Here $|\psi(t)\rangle$ is again our chosen trajectory and $|\psi_{\rm Q}(t)\rangle$ is the state obtained from iTEBD~\cite{Vidal07} simulations. 

    We show the results of one such simulation in Fig.~\ref{fig:comp}c, where we compare the fidelity density of the three approaches in driving the state around a circular trajectory defined by Eq.~\eqref{Eq:circle}. 
    We observe that the leakage minimization approach outperforms both CD approaches. 
    We find that the best performance of the leakage-based approach consistently holds  across several tested trajectories, although in some cases the ground state CD approach offers comparable fidelity. 
    Nevertheless, given the additional requirement of finding a parent Hamiltonian, there is no advantage to using it instead of the leakage approach.
    The trace CD approach is by far the worst of the three. 
    While in the example shown in Fig.~\ref{fig:comp}(c) it is able to follow the trajectory with some success, for trajectories farther from the scar trajectory, the trace CD approach fails completely~(see Appendix~\ref{app:trajectories}). 
    Hence, in the remainder of this work we focus only on the leakage-based approach that provides the optimal driving parameters for a certain fixed trajectory. 

    The example we have chosen can also be viewed as related to scar stabilization through periodic driving, that was recently explored both experimentally and theoretically~\cite{Dolev20,Nishad21}.
    Essentially we are able to construct Floquet scars in an arbitrary model, provided a good solution exists within the constraints --- the choice of MPS manifold $\Psi$ and control operators $A_\eta$. 
    In the particular case of the PXP model this yields a stabilization of the PXP scar by means of weak periodic driving.
    
    Our construction of the leakage-based approach to optimal control suggests a natural next step: to optimize over all possible trajectories connecting given final and initial states, $|\psi(0)\rangle$ and $|\psi(\tau)\rangle$. 
    This is equivalent to the common formulation of the state preparation problem. 
    While a complete solution to this problem is beyond the scope of the present work, in the next Section we visualize the cost function (leakage) landscape in a simple case and illustrate trajectory optimization over the variational parameters.  
    
    \section{Leakage landscape and entanglement steering}
    \label{sec:landscape}
    In this section we go beyond the local leakage minimization framework constructed in Sec.~\ref{sec:control}. 
    First, we study how the efficiency of steering of the quantum system depends on the chosen direction of the unitary evolution in the tangent space of the MPS manifold specified by $\partial_t |\psi\rangle$. 
    This leads to the definition of the optimal direction, where the rescaled leakage $\Delta$ is minimal.
    We then apply this to the generalized PXP model in order to find a trajectory with minimal leakage. 
    In addition, we demonstrate the optimization of the scarred trajectory in the generalized PXP model that allows us to control the amount of entanglement generated during the evolution without degrading the fidelity. 

    \begin{figure*}
        \centering
        \includegraphics[width=\textwidth]{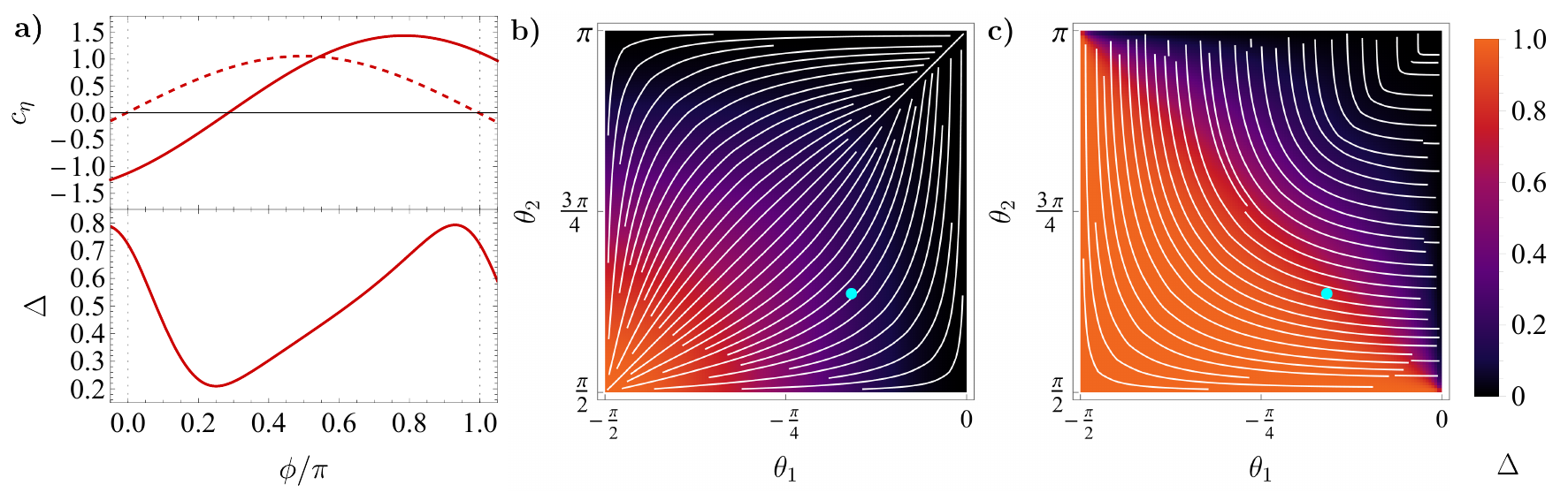}
        \caption{
            (a) The dependence of driving parameters $c_{1,2}$ and rescaled leakage $\Delta$ with respect to the desired direction of travel $\phi$ at the point $(\theta_1,\theta_2)=(-0.5,2.0)$ (marked by a light blue point in panels (b) and (c)). 
            The lowest value of $\Delta$ determines the optimal direction.
            (b) The heat map shows the value of minimal leakage, and contour lines indicate the optimal direction in the parameter space of the MPS. 
            Panel (c) illustrates the maximal possible magnitude of the leakage along with the direction where it is highest. 
        }
        \label{fig:leakSD}
    \end{figure*}
    
    \subsection{Optimal direction and minimal leakage}
    
    When a trajectory is fixed, the leakage minimization discussed above provides a specific solution for the optimal  $A_\text{opt}$ that generates dynamics closest to the \emph{particular} target direction, $\partial_t |\psi(t)\rangle$. 
    In addition, the dimensionless leakage $\Delta^2$ introduced in Eq.~\eqref{eq:leakAngle} allows us to quantify how well we are able to steer the quantum system along the chosen tangent vector. 
    Given such a solution to the optimal control problem for a \emph{fixed} tangent vector, it is natural to consider how the generator of the unitary evolution and resulting dimensionless leakage $\Delta^2$ depend on the chosen direction within the tangent plane of the MPS manifold. 
    
    Since dimensionless leakage $\Delta^2$ is a continuous function of the direction within the tangent plane bounded between 0 and 1, it achieves a minimum at a certain point. 
    We call the direction of evolution that minimizes $\Delta^2$ the \emph{optimal direction}, since it intuitively corresponds to the direction where the steering of the quantum state is most efficiently implemented using the allowed set of control parameters. 
    
    Having discussed the existence of an optimal direction with minimal leakage, we illustrate these notions using the generalized PXP model. 
    The chosen MPS manifold is parametrized by two angles $\theta_1$ and $\theta_2$, with its tangent plane being spanned by two vectors $\partial_{\theta_{1,2}}|\psi\rangle$ at each point.
    In other words, the different directions of evolution in the tangent plane of the MPS manifold correspond to the different directions of the trajectory in the $(\theta_{1},\theta_{2})$ plane passing through a particular point. 
    For instance, Fig.~\ref{fig:leakSD}(a) illustrates these concepts for the particular point $(\theta_1,\theta_2)=(-0.5,2.0)$ of the variational manifold. 
    The direction of the trajectory $\phi$ defines values of derivatives as $\theta'_1(t)=r\cos\phi$ and $\theta'_2(t)=r\sin\phi$. 
    We note that the velocity in the parameter space $r$ has only a trivial effect of proportionally rescaling the driving parameters $c_\eta\propto r$, see e.g.\ Eq.~(\ref{eq:leakC}). 
    From this it is easy to see that the rescaled leakage $\Delta$ is independent of this velocity. 
    
    In the top panel of Fig.~\ref{fig:leakSD}(a) we observe that the optimal values of parameters $c_{1,2}$ that specify the operator $A$ strongly depend on the  angle $\phi$ that specifies the direction of the evolution in the tangent plane. 
    The bottom panel shows that $\Delta^2$ changes depending on the direction, assuming a minimal value for $\phi\approx \pi/4$, which corresponds to the optimal direction of evolution at the chosen point of the variational manifold.
    This represents the direction in which we can move with the lowest leakage by using an operator from the chosen set of driving operators $A$.
    Similarly, we can also identify the worst direction where leakage is largest. 
    Note that the optimal and worst directions are typically not orthogonal. 
    Furthermore, leakage stays invariant under replacement  $\phi\to \phi+\pi$, as moving backwards or forward is equivalent in terms of leakage (the expression for the optimal driving parameters changes sign when direction of evolution is flipped).
    
    After illustrating the concepts of optimal direction and associated minimal leakage at a particular point of the variational MPS manifold, we continue with the study of these concepts throughout the entire manifold. 
    Importantly, looking at the best (worst) direction and corresponding leakage should provide us with sufficient information to chart a relatively good course between specific states in the MPS manifold. 
    Figure~\ref{fig:leakSD}(b)-(c) shows such information for the particular example of generalized PXP model and the two-parameter MPS state defined in Sec.~\ref{sec:model}. 
    
    Focusing on the optimal direction and associated minimal value of the leakage in Fig.~\ref{fig:leakSD}(b), we observe that minimal leakage is small far away from the point $(\theta_1,\theta_2)=(-\pi/2,\pi/2)$ and increases upon approaching this point.
    Such behavior of the leakage may be attributed to the fact that the point $(-\pi/2,\pi/2)$ corresponds superposition of two macroscopically distinct states as discussed in Sec.~\ref{sec:model}. 
    The value of minimal leakage approaches one in vicinity of this point, suggesting that \emph{all} directions of unitary evolution are nearly orthogonal to the MPS tangent plane, irrespective of the chosen values of the driving parameters $c_{1,2}$ in $A$. 
    This implies that one should avoid that area of the MPS manifold, as the available controls become inefficient. 
    
    Additional information is revealed by comparing Fig.~\ref{fig:leakSD}(b) with Fig.~\ref{fig:leakSD}(c) that shows the worst possible driving direction and associated leakage. 
    At the top and right edges of the chosen parameter region, both minimal and maximal leakage are small, indicating that in this area one can very efficiently manipulate the state of the system with the available controls. 
    This can be explained by the fact that line $\theta_{1}= 0$ ($\theta_2=\pi$)  correspond to the product state of the system where all odd (even) sites of the chain are in the $\downarrow$ state. 
    Thus moving along along two segments $(-\pi/2,\pi)\to(0,\pi)\to(0,\pi/2)$ one can very efficiently drive the system between $|Z_2\rangle$ and $|Z'_2\rangle$ product states as $|Z_2\rangle\to|\hspace{-1mm}\downarrow\downarrow\downarrow\dots\rangle\to|Z'_2\rangle$ without creating any entanglement.   

    In contrast to the behavior discussed above, the center of the parameter region, $(\theta_1,\theta_2)\approx(-\pi/4,3\pi/4)$ is characterized by strong anisotropy of $\Delta$ with respect to the chosen direction of the trajectory. 
    While moving along one of the diagonal directions can be performed reasonably efficiently, the motion along the orthogonal diagonal causes large leakage that approaches one. 
    This implies that controlling the quantum many-body state in that parameter region near the point $(\theta_1,\theta_2)\approx(-\pi/4,3\pi/4)$ is more challenging.
    At the same time, in this parameter region the MPS has a non-trivial pattern of entanglement. 
    Thus, in order to illustrate entanglement steering~\cite{Dolev20} we optimize trajectories passing through this region, which is presented in the following Section. 
    
    \subsection{Entanglement steering and trajectory optimization}
    \label{sec:scar}
    In order to illustrate the trajectory optimization, we introduce a two parameter family deformation of the circle trajectory considered in Fig.~\ref{fig:comp} from Sec.~\ref{sec:control}.  This family is defined as 
    \begin{equation}
        \label{eq:angles}
        \begin{split}
            &\theta_1(t)=\frac{\pi}{2}\left[d(t)\sin\pi t-1\right], 
            \\
            &\theta_2(t)=\frac{\pi}{2}\left[d(t)\cos\pi t+1\right],
        \end{split}
    \end{equation}
    \begin{equation}
        d(t)=1-\varepsilon_1(1-\cos2\pi t)-\varepsilon_2(1-\cos4\pi t)), 
        \label{eq:deform}
    \end{equation}
    where deformation parameters $\varepsilon_1$ and $\varepsilon_2$ modify the local radius of the circle. 
    For $\varepsilon_{1,2}=0$ the trajectory is the undeformed segment of the circle. The $\varepsilon_1$ parameter can be used to determine the intersection point with the diagonal line $\theta_1=\theta_2-\pi$. 
    By pushing the point of intersection with the diagonal towards the bottom left corner $(-\pi/2,\pi/2)$ one can increase the maximal amount of bipartite entanglement entropy $S_t=-{\rm tr}{\rho_A(t)\log\rho_A(t)}$ reached in the dynamics along the trajectory, where $\rho_A$ is the reduced density matrix of one half of the system. 
    The parameter $\varepsilon_2$ deforms the trajectory between the times $t=\{0,\tau/2,\tau\}$ and can be used to optimize the trajectory once $\varepsilon_1$ has been fixed. 
    
    \begin{figure}[tb]
        \includegraphics[width=\columnwidth]{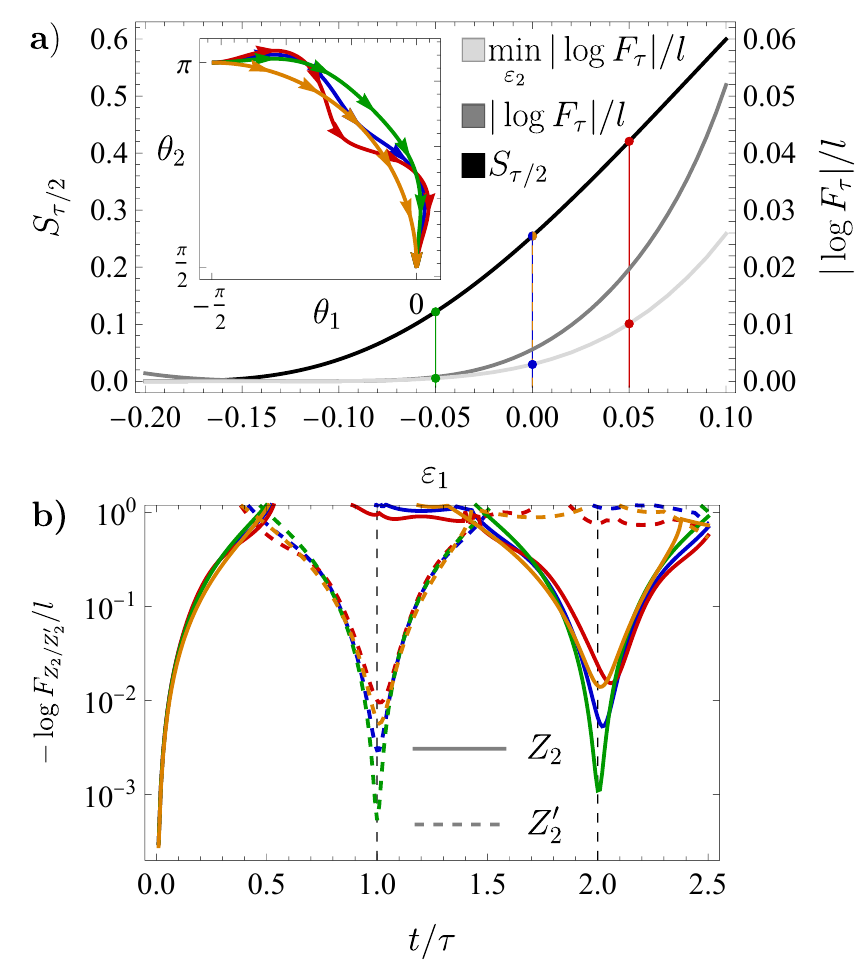}
        \caption{
            (a) The dependence of entanglement entropy at $t=\tau/2$ and fidelity at $t=\tau$ with respect to a single trajectory deformation parameter $\varepsilon_1$ at $\varepsilon_2=0$. 
            The inset shows example optimized trajectories at $\varepsilon_1=-0.05,0,0.05$ and $\varepsilon_2=-0.02,-0.05,-0.08$ (green, blue and red respectively), the increase in fidelity is related to these trajectories avoiding the high-leakage areas depicted in Figure 3. 
            We observe that at $\varepsilon=-0.05$ we can obtain relatively high fidelity when driving the state from a product state to an entangled one at $t=\tau/2$ and back.
            In panel \textbf{b)} we show the overlaps with the $Z_2/Z'_2$ states for these trajectories (color-coded as in panel \textbf{a)}) as well as the scar trajectory (orange).
        }
        \label{fig:scar2}
    \end{figure}
    
    The parametrization~\eqref{eq:angles}-\eqref{eq:deform} makes the trajectory optimization problem finite-dimensional. 
    While this is not guaranteed to provide the best possible solution for the trajectory connecting the $|Z_2\rangle$ state to the given point on the diagonal, it allows for an efficient optimization illustrating the capabilities of our approach for controlling entanglement.
    First, in Fig.~\ref{fig:scar2}(a) we show how the entanglement at the middle point of trajectory depends on the parameter $\varepsilon_1$. 
    Negative values of $\varepsilon_1$ push trajectory towards the region of small entanglement, whereas positive $\varepsilon_1\approx 0.1$ brings entanglement close to the value of $\log 2$. 
    Since the value of $\varepsilon_1$ sets the amount of entanglement to be achieved within the course of dynamics, we illustrate the trajectory optimization for three different values of this parameter, $\varepsilon_1= -0.05$, $0$, and $0.05$.
   
    The value of $\varepsilon_1= -0.05$ corresponds to the deformation of the original scar trajectory that decreases entanglement in the course of quantum evolution [green line and point in Fig.~\ref{fig:scar2}(a)]. 
    Since the trajectory passes relatively far from the dangerous high leakage region, the optimization over parameter $\varepsilon_2$ does not improve fidelity visibly. 
    Next, we consider the value of $\varepsilon_1= 0$ that is close to the original scar trajectory and corresponds to approximately twice larger value of the bipartite entanglement reached in the evolution. 
    The optimization of trajectory using  parameter $\varepsilon_2$ allows to improve fidelity of revivals as is shown by the blue dot in Fig.~\ref{fig:scar2}(a). 
    Not surprisingly, the inset illustrates that the effect of the optimization is to push the trajectory towards the lower leakage regions at the top and right edges of the parameter region, which agrees with the intuition provided by Fig.~\ref{fig:leakSD}. 
    
    Finally, at $\varepsilon_1= 0.05$ we reach even higher entanglement, which comes with degraded fidelity. 
    Optimization over $\varepsilon_2$ increases associated fidelity, that at the first revival in Fig.~\ref{fig:scar2}(b) reaches a better value compared to the un-optimized trajectory with $\varepsilon_2=0$ at fixed $\varepsilon_1$. 
    The general trend that emerges from our optimization is that increasing the amount of entanglement encountered during evolution leads to lower fidelity. 
    This may be attributed to the very limited amount of controls available in the present example: we only allow to change the values of Rabi frequency on two sub-lattices controlled by parameters $c_{1,2}$ in Eq.~\eqref{eq:pxp}. 
    We expect that one should be able to increase the fidelity with which we can reach the higher entangled states by adding additional control parameters, thus allowing more efficient exploration of larger parts of the MPS manifold.  

    \section{Floquet scars in the transverse and longitudinal field Ising model} 
    \label{sec:ising}
    
    We now apply our approach to the transverse and longitudinal field Ising model (TLFIM) defined by the following Hamiltonian,
    \begin{equation}
        \label{eq:ising}
        A_{\rm TLFIM}=\sum_i\left(J\sigma^z_i\sigma^z_{i+1}+h_z\sigma^z_i+h_x\sigma^x_i\right).
    \end{equation}
    We limit the dynamics to a simple bond dimension 2 manifold defined by
    \begin{equation}
        \label{eq:isingmps}
        M=\begin{pmatrix}
            \cos\frak{d}\cos\frak{b} e^{i\frak{a}/2}|\hspace{-1mm}\uparrow\rangle & \cos\frak{d}\sin\frak{b} e^{-i\frak{a}/2}|\hspace{-1mm}\uparrow\rangle \\
            \sin\frak{d}\sin\frak{b} e^{i(\frak{c}-\frak{a}/2)}|\hspace{-1mm}\downarrow\rangle & \sin\frak{d}\cos\frak{b} e^{i(\frak{c}+\frak{a}/2)}|\hspace{-1mm}\downarrow\rangle
        \end{pmatrix},
    \end{equation}
    that was proposed in Ref.~\cite{Michailidis2020}.
    Unlike in the PXP model, we consider states with 1-site translational invariance. 
    
        \begin{figure*}[t]
        \centering
        \includegraphics[width=1.0\linewidth]{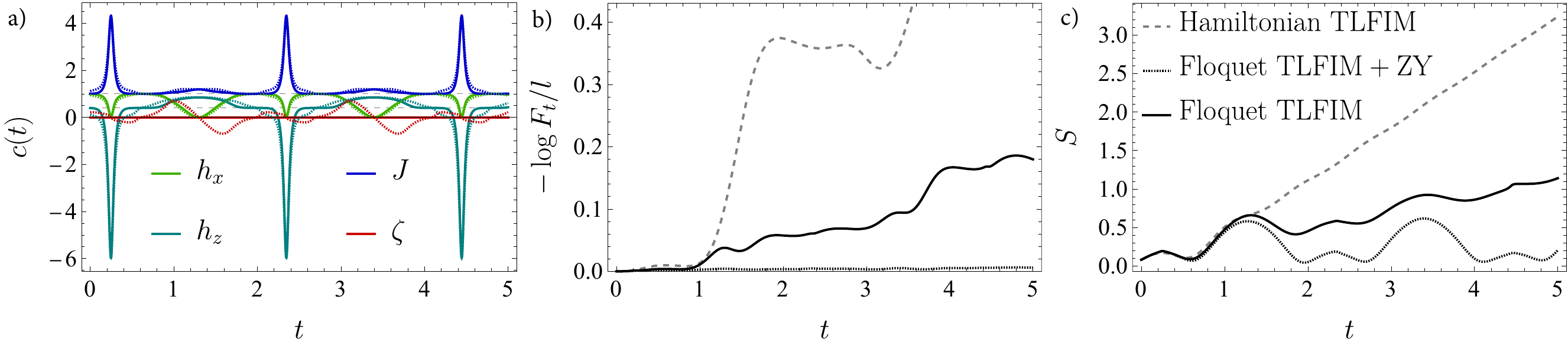}
        \caption{
            {(a) The time dependence of the optimal values of the parameters for the TLFIM (full lines) and the perturbed TLFIM (dotted lines). 
            The dashed lines show the values of the parameters in the time-independent case used to generated the trajectory. 
            (b) The logarithm of fidelity density obtained from iTEBD simulations~\cite{Vidal07}, where $F_t=|\langle\psi(t)|\psi_{\rm Q}(t)\rangle|^2$, measures how close exact unitary evolution $|\psi_{\rm Q}(t)\rangle$ is to the desired MPS state, $|\psi(t)\rangle$. 
            We observe a significant improvement of fidelity for Floquet TLFIM, and nearly perfect fidelity for TLFIM supplemented by additional terms (denoted as TLFIM+ZY).
            (c)  In line with the behavior of fidelity, the bipartite entanglement grows much slower for Floquet TLFIM, and its growth is hardly noticeable for extended Floquet driving. 
            All data shown were computed with $\chi=64$ and $\delta t=10^{-5}$. 
        }}
        \label{fig:ising}
    \end{figure*}
    
    Using a TDVP projection of the dynamics on the MPS state~(\ref{eq:isingmps}),  Ref.~\cite{Michailidis2020} found a periodic trajectory for fixed values of couplings $J=1$, $h_z=0.4$, and $h_x=1$. 
    This trajectory, for instance can be identified by the particular value of MPS parameters, $(\frak{a},\frak{b},\frak{c},\frak{d})=(0.2607,0.9,4.888,0.4308)$ and has a period of $t_0=2.097$~\cite{Michailidis2020} (see Appendix~\ref{app:ising} for details on the trajectory). 
    However, TDVP dynamics on this trajectory encounters regions with large leakage, thus not giving rise to significant fidelity revivals, nor to non-ergodic ``scarred'' eigenstates in the spectrum of Ising model~(\ref{eq:ising}). 
    Below we use the optimal steering approach to find the optimal time-dependent values of couplings in the Ising model for steering the system along the periodic trajectory. 
    
    First we treat three couplings in the Ising model, (\ref{eq:ising}) as  variational parameters $c_{1,2,3}$. Using Eqs.~(\ref{eq:D})-(\ref{Eq:c-general-leakage}), we  obtain values of the Ising model parameters $J$, $h_{z}$, and $h_x$ which are shown in Fig.~\ref{fig:ising}(a) by solid lines. 
    Since we consider a closed trajectory, the time dependent values of couplings obtained from our steering approach effectively define a Floquet model. 
    We  observe that the values of couplings agree with their ``static'' values for particular sections of trajectory. 
    This can be attributed to the overall low leakage in those areas, hence suggesting that constant values of couplings $J=1$, $h_z=0.4$, and $h_x=1$ are nearly optimal in that region. 
    Outside of the regions where couplings are close to their static values, the optimization gives large peaks in the driving parameters. 
    These peaks are the result of a large derivative of the state at that point along the trajectory and can be readily eliminated by modifying the velocity for a particular region of the trajectory. 
    
    Comparing exact unitary dynamics generated by the operator $A_{\rm TLFIM}$  with constant (Hamiltonian TLFIM) and time dependent couplings (Floquet TLFIM), we observe a significant improvement in fidelity in Fig.~\ref{fig:ising}(b). 
    Since the motion is occuring on the periodic trajectory in the MPS manifold, the system will exhibit periodic revivals of the fidelity of the initial state, characteristic of quantum many-body scars. 
    We note that the fidelity of the revival can be further improved by either modifying the trajectory, as we had done in the previous section, or by adding additional driving parameters which would allow us to follow the fixed trajectory with greater precision. 
    
    Among all possible two-site operators, the best additional term to the operator $A_{\rm TLFIM}$ turns out to be 
    \begin{equation}
        \delta A=\frac12\zeta\sum_i\left(\sigma^z_i\sigma^y_{i+1}+\sigma^y_i\sigma^z_{i+1}\right),
    \end{equation}
    such that the full generator of unitary dynamics becomes $A=A_{\rm TLFIM}+\delta A$. 
    As can be seen in Fig.~\ref{fig:ising}(a) such an addition leads to a minor change in the other driving parameters, and at the same time yielding a non-trivial time dependence for $\zeta$, whose average across the entire trajectory is exactly zero. 
    In Fig.~\ref{fig:ising}(b) we see that this term allows us to reach near perfect fidelity on the desired trajectory. 
    
    Finally, in Fig.~\ref{fig:ising}(c), we study the dynamics of the bipartite entanglement entropy. 
    In the case of Hamiltonian dynamics (constant couplings) we observe generic linear growth of entanglement entropy. 
    Conversely, our attempt to stabilize the dynamics on the MPS trajectory using the Floquet TLFIM model alone immediately yields a notable slowdown in the rate of growth of the entanglement entropy [solid line in Fig.~\ref{fig:ising}(c)]. 
    Finally, the Floquet dynamics generated by the operator $A_{\rm TLFIM}+\delta A$ nearly completely halts entanglement growth, leading to periodic oscillation of entanglement akin to those observed in the PXP model~\cite{Michailidis2020}.
    
    \section{Discussion}
    \label{sec:discussion}
    In this work we considered the problem of steering a quantum many-body system along a certain path in the MPS manifold.
    We introduced the leakage minimization approach as the most promising and efficient general solution for this problem. 
    In order to illustrate this approach, we considered the generalized PXP model where we allow for time-dependent control of the Rabi frequency on even and odd sites of the chain.
    This additional control was used to improve the fidelity of revivals in the unitary dynamics generated by our time-dependent Hamiltonian. 
    In addition, we demonstrated how one can control the amount of quantum entanglement encountered in such unitary dynamics, and showed that one may perform entanglement steering~\cite{Dolev20} using a limited set of controls. 
    Finally, we applied our approach to the non-integrable transverse and longitudinal field Ising model. 
    We found the time dependent values of couplings in this model that give rise to fidelity revivals characteristic of quantum many-body scars. 
    
    Earlier works used MPS encoding of the quantum state for more efficient numerical evaluation of the cost function in the optimal control approaches~\cite{Doria11}. 
    In contrast, our method finds the optimal controls for steering the system along a given direction in the MPS tangent space, and is capable of determining efficient and inefficient directions of propagation given a certain set of control parameters. 
    Thus, our approach intrinsically relies on the MPS representation of the quantum many-body wave function to map the quantum control problem to a classical one in an efficient way.
    Although we focused on an analytical demonstration of this method using a small number of parameters and a bond dimension two MPS state, it can be straightforwardly extended to the case of larger bond dimensions and more control parameters by performing the calculations numerically.
    While our approach is naturally formulated in the thermodynamic limit, similarly to the TDVP method~\cite{Haegeman}, thus providing an efficient solution for the problem of locally steering of quantum many-body system dynamics, the generalization to finite size systems is also straightforward.
    
    Moreover, our approach can be extended beyond the MPS representation to other variational manifolds of states. 
	The framework proposed in our work remains efficient and avoids complexity related to the full many-body Hilbert space, provided that the variational manifold allows for an efficient parametrization of the tangent vectors and for an efficient calculation of two-point correlation functions. 
	Prospective extensions of our framework are provided by two-dimensional projected entangled pair states (PEPS), and also variational states where observables can be calculated with Monte Carlo sampling. 
	An example of a specific extension of our approach could be the optimal state preparation of spin liquid type resonant valence bond states in two-dimensional Rydberg arrays, recently implemented experimentally~\cite{Seme21} and studied theoretically using a few-parameters PEPS ansatz~\cite{Pichler}. 
    
    Pairing the efficient solution of the local steering problem constructed in this work with the global trajectory optimization within a given MPS variational manifold is expected to lead to a new approach to the quantum control problem (see Appendix~\ref{app:euler-lagrange} for the formulation of the trajectory optimization problem). 
    Although MPS representation is limited in its capabilities of representing strongly entangled states, this restriction can be practically avoided by choosing a sufficiently large bond dimension. 
	For instance, numerical MPS-based trajectory optimizations of the state preparation across superfluid-insulator phase transition in the Bose Hubbard model is reported to be converged for a bond dimension of $\chi=200$~\cite{Jesper21}.
    In addition, many interesting states that have a non-trivial global entanglement pattern, such as the celebrated Greenberger-Horne-Zeilinger state, are naturally encoded as MPS of a low bond dimension.
    Hence, it is interesting to understand how our framework will perform in preparing the globally entangled states and facilitate the state preparation in quantum simulators.
    
    Finally, the problem of the steering of quantum states has potentially fruitful connections with quasi-adiabatic continuity results~\cite{Hastings04,Hastings10}. 
    The latter guarantee that two many-body states, connected by a trajectory along which the system remains gapped, can be transformed into each other by an evolution with a fictitious quasilocal Hamiltonian over a finite time. 
    This Hamiltonian is expressed via spectral properties and eigenstates, and therefore is generally difficult to implements in experiments. 
    Our approach provides a practical recipe to constructing approximate quasi-adiabatic evolution protocols, provided a finite number of controls. 
    Theorems of Refs.~\cite{Hastings04,Hastings10} ensure that if a set of controls is sufficiently large, the desired state can be prepared with desired accuracy.
    In the future, it would be interesting to investigate the minimum set of controls required to drive a quantum state between two quantum states. 
    
    On a more practical note, our application of the leakage minimization control framework to the generalized PXP model reveals a new perspective on quantum many-body scars~\cite{Moudgalya21,Serbyn21}.
    Our results show that a weak time-dependent modulation of Rabi frequencies on different sub-lattices allows one to make the characteristic quantum scarred dynamics more coherent.
    Such a modulation of Rabi frequencies effectively leads to a Floquet model with scars characterized by a smooth dependence of Hamiltonian parameters. 
    Although Floquet scars in models with pulsed driving received significant attention~\cite{PhysRevB.102.224309,Mukherjee2020,Mizuta2020,Nishad21}, our work invites the study of Floquet many-body scars with continuous driving.
    
    More broadly, our approach may be used to construct new models with Floquet scars, as we explicitly demonstrated using the Floquet TLFIM. 
    We expect that this will enable realization of quantum many-body scars using other quantum simulator platforms, and can reveal additional insights into a general mechanism for the appearance of scars~\cite{Moudgalya21,Serbyn21}. 
    Natural candidate models for the construction of Floquet scars include the PXP model with time dependent chemical potential~\cite{Dolev20,Nishad21}, and other lattice models that have a known frustration-free MPS ground state, such as AKLT spin chains~\cite{Bernevig2017,BernevigEnt}, Hubbard type models~\cite{MoudgalyaHubbard} and others~\cite{MoudgalyaFendley} that are also known to have quantum many-body scars. 

    \section*{Acknowledgments} 
    \label{sec:acknowledgements}
    
    We thank A.~A.~Michailidis for insightful discussions. 
    M.~L. and M.~S. acknowledge support from the European Research Council (ERC) under the European Union’s Horizon 2020 research and innovation programme (grant agreement No.~850899). 
    D.~A. was supported by the European Research Council (ERC) under the European Union’s Horizon 2020 research and innovation programme (grant agreement No.~864597) and by the Swiss National Science Foundation.
    The iTEBD simulations were performed using the ITensor library~\cite{itensor}. 

    \appendix    
    \section{CD approach}
    \label{app:cd}
    
    \subsection{Trace minimization}
    \label{app:trace}
    
    In this section we will briefly discuss the calculation of the trace CD approach in general and in the PXP model.
    \subsubsection{General expression for the cost function}
    We begin with the expression for the cost function
    \begin{equation}
        S(t)=\tr\left[G(t)^2\right].
    \end{equation}
    Typically we assume a translational invariance with possibly large but finite unit cell (this is required for calculations in the thermodynamic limit $l\to\infty$). 
    In such cases it is possible to expand the expression for the cost function,
    \begin{equation}
        S(t)=\tr\left[\sum_{i,j}G_i(t)G_j(t)\right],
    \end{equation}
    where we introduced the density $G_j(t)$ as 
    \begin{equation}\label{Eq:G-local}
        G_j(t)=\partial_tH_j(t)+i\sum_{j'}[A_{j'}(t),H_j(t)].
    \end{equation}
    Provided both $A_j(t)$ and $H_j(t)$ are local operators, the sum over $j'$ has a finite number of non-zero terms. 
    There will always exist an integer $l_+$ such that $[A_{j+l}(t),H_j(t)]=0;\forall~l\ge l_+$. 
    Similarly, there will be an integer $l_-$ such that $[A_{j-l}(t),H_j(t)]=0;\forall~l\ge l_-$. 
    Hence we can rewrite Eq.~(\ref{Eq:G-local}) as
    \begin{equation}
    \label{G-simplified}
        G_j(t)=\partial_tH_j(t)+i\sum_{j'=-l_-+1}^{l_+-1}[A_{j+j'}(t),H_j(t)],
    \end{equation}
    where $l_+\le l_H$ and $l_-\le l_A$, where $l_A$ and $l_H$ are the supports of $A_j$ and $H_j$ respectively~(the densities are taken over translationally invariant unit cells). 
    From here it follows that $G_j(t)$ is a local operator with support on most $2l_A+l_H-2$ consecutive sites.
    
    Let us now define a density of the cost function as
    \begin{equation}
        \label{seq:1}
        S_{i,j}(t)=\tr\left[G_{i,j}(t)\right]
        \quad
        G_{i,j}(t)=G_i(t)G_j(t). 
    \end{equation}
    Using this definition, the full cost function can be expressed as
    \begin{equation}
        \label{seq:2}
        S(t)=\frac{l}{l_U}\sum_{i=1}^{l_U}\sum_{j}S_{i,i+j}(t),
    \end{equation}
    where $l_U$ is the size of the unit cell (i.e. we have translational invariance with respect to shifting the system by $l_U$) and $l$ is the size of our system. 
    Finally, since $A_j$ is linear in $c_\eta(t)$ it follows that $S(t)$ will be a quadratic polynomial in $c_\eta(t)$. 
    We can then write the cost function in a generic form as
    \begin{equation}
        S(t)=\sum_{\eta,\eta'}C_{\eta,\eta'}(t)c_\eta(t) c_{\eta'}(t)+\sum_{\eta}C_{\eta}(t)c_{\eta}(t)+C(t).
    \end{equation}
    Here $C_{\eta,\eta'}(t)$, $C_\eta(t)$ and $C(t)$ do not depend on $c_\eta(t)$. 
    Clearly the value of the optimal parameters $c_\eta$ will not depend on $C(t)$, so we can simply omit the contribution from terms of the form $\tr(\partial_tH_i(t)\partial_tH_j(t))$. 
    This is practical, since it will allow us to truncate the sum in $j$ in Eq.~\eqref{seq:2} to some finite range, which we can do since $A$ and $H$ are both Hermitian. 
    For $C_{\eta,\eta'}$ this can be seen by looking at terms of the form $\tr\left([A_{i+j},H_{i}][A_{i'+j'},H_{i'}]\right)$ that appear when in Eq.~(\ref{seq:1}) one takes two terms that both contain commutators. 
    Such contributions will cancel out with $\tr\left([A_{i'+j'},H_{i'}][A_{i+j},H_{i}]\right)$ when the two commutators commute (for instance when the supports of the resulting local operators do not overlap). 
    A similar observation can be made for $C_\eta$ with $\tr\left((\partial_tH_i(t))[A_{i'+j'},H_{i'}]\right)$ and $\tr\left((\partial_tH_{i'}(t))[A_{i+j},H_{i}]\right)$.
    From this, we can see that it is sufficient to evaluate the following expression
    \begin{equation}
        S=\sum_{i=1}^{l_U}\sum_{j=-l_+-l_-+2}^{l_++l_--2}\tr\left[S_{i,i+j}\right]+{\rm constant},
    \end{equation}
    where we ignore parts of the term constant in $c_\eta$, that does not impact minimization over $c_\eta$. 
    
    \subsubsection{Calculation for the PXP model}
    \label{app:trace-calc}

    When applying the above approach to the PXP model we only need to evaluate the following expression
    \begin{equation}
        S=\frac{l}{2}\sum_{i=1}^{2}\sum_{j=-2}^{2}S_{i,i+j}+{\rm constant}.
    \end{equation}
    An important detail here is that the trace in $S_{i,j}(t)$ [defined in Eq.~\eqref{seq:1}] is taken only with respect to the constrained subspace (i.e.\ excluding states where two neighboring sites are occupied). 
    As a result if one works with the basis of Pauli matrices, the $\sigma^x$ and $\sigma^y$ matrices remain traceless while strings of operators containing $\sigma^z$ may have non-zero trace in this projected subspace. 
    To compute the trace in the constrained subspace, let us expand the trace as sum over all basis vectors (in the $\sigma^z$ computational basis) in the subspace
    \begin{equation}
        S_{i,j}=\tr\left[G_i(t)G_j(t)\right]=\sum_{\{|\xi\rangle\}}\langle\xi|G_i(t)G_j(t)|\xi\rangle.
    \end{equation}
    Additionally, we can expand the operators themselves in terms of Pauli strings
    \begin{equation}
        G_i(t)G_j(t)=\sum_{\underline{\alpha}}g_{\underline{\alpha}}\bigotimes_{i}\sigma_i^{\alpha_i},
    \end{equation}
    where $\underline{\alpha}=\{\alpha_1,\alpha_2,\dots,\alpha_l\}$ and $\alpha_i\in{0,x,y,z}$.
    Clearly, any strings containing $\sigma^x$ or $\sigma^y$ will not contribute, since those are purely off-diagonal in the computational basis. 
    What is left is to compute an overlap of some string of $\sigma^z$ or $\id$ operators with all basis states (trace)
    \begin{equation}
        t_{\underline{\alpha}}=\frac{1}{D}\sum_{\{|\xi\rangle\}}\langle\xi|\bigotimes_{i}\sigma_i^{\alpha_i}|\xi\rangle,
    \end{equation}
    in the thermodynamic limit $l\to\infty$. 
    Here $D$ is the dimension of the constrained subspace that also depends on $l$. 
    To do this, let us first look at a chain of length $l$ with periodic boundary conditions.
    In this case, one can construct the sum of all basis states spanning the constrained subspace using MPS formalism as 
    \begin{equation}
        \sum_{\{|\xi\rangle\}}|\xi\rangle=\tr\left[T_0^l\right],\hspace{1cm}T_0=\begin{pmatrix}|\hspace{-1mm}\downarrow\rangle & |\hspace{-1mm}\downarrow\rangle \\ |\hspace{-1mm}\uparrow\rangle & 0\end{pmatrix}.
    \end{equation}
    Applying $\sigma^z_i$ to such an MPS state a specific site $i$ simply requires replacing the $i$-th transfer matrix $T_0$ with $T_z$
    \begin{equation}
        T_z=\begin{pmatrix}-|\hspace{-1mm}\downarrow\rangle & -|\hspace{-1mm}\downarrow\rangle \\ |\hspace{-1mm}\uparrow\rangle & 0\end{pmatrix}.
    \end{equation}
    Obtaining an expectation value for a string of $\id$ and $\sigma^z$ matrices is then simply expressed as
    \begin{equation}
        t_{\underline{\alpha}}=\frac{1}{D}\left(\tr\left[T_0^l\right]\right)^\dagger\tr\left[\prod_{i=1}^lT_{\alpha_i}\right].
    \end{equation}
    The scalar factors in front of each basis state $|\xi\rangle$ are simply $(-1)^{n_\downarrow}$, where $n_\downarrow$ is the number of $|\hspace{-1mm}\downarrow\rangle$ in the state $|\xi\rangle$.
    This allows us to further simplify the expression into
    \begin{equation}
        t_{\underline{\alpha}}=\frac{1}{D}\tr\left[\prod_{i=1}^l\widetilde{T}_{\alpha_i}\right].
    \end{equation}
    Here we introduced new transfer matrices as
    \begin{equation}
        \tilde{T}_0=\begin{pmatrix}1&1\\1&0\end{pmatrix},\hspace{1cm}\widetilde{T}_z=\begin{pmatrix}-1&-1\\1&0\end{pmatrix},
    \end{equation}
    which now already include the overlap with $\langle\psi|$ and can be obtained from $T_\alpha$ by simply replacing $|\hspace{-1.3mm}\uparrow\rangle\to1$ and $|\hspace{-1mm}\downarrow\rangle\to1$. 
    
    Taking the thermodynamic limit of this expression in our case is relatively straightforward, since all our operators are local (i.e. $\alpha_i=0$ outside some fixed interval). 
    When that is the case, we can rewrite the expression as
    \begin{equation}
        t_{\underline{\alpha}}=\lim_{l\to\infty}\frac{1}{D}\tr\left[\widetilde{T}_0^{l_{\rm left}}\left(\prod_{i\in{\rm center}}\widetilde{T}_{\alpha_i}\right)\widetilde{T}_0^{l_{\rm right}}\right],
    \end{equation}
    where $l_{\rm left}$ and $l_{\rm right}$ are the length of the strings of identities to the left/right of the nontrivial local operator respectively. 
    We also require $l_{\rm left}$ and $l_{\rm right}$ to both diverge as we take $l\to\infty$, which we can always choose due to translational invariance in the system. 
    When $T$ has a single dominant eigenvalue, which is also the case here, we have
    \begin{equation}
        \lim_{n\to\infty}\widetilde{T}_0^n=\lim_{n\to\infty}\lambda_{\rm max}^n|e_{\rm max}\rangle\langle e_{\rm max}|,
    \end{equation}
    where $\lambda_{\rm max}$ is the largest (by absolute value) eigenvalue and $|e_{\rm max}\rangle$ is the corresponding eigenvector. 
    Note that the left and right eigenvectors are the same in this case.
    Replacing this in the trace we obtain
    \begin{equation}
        t_{\underline{\alpha}}=\lim_{l\to\infty}\frac{1}{D}\lambda_{\rm max}^{l-l_{\rm center}}\langle e_{\rm max}|\prod_{i\in{\rm center}}\widetilde{T}_{\alpha_i}|e_{\rm max}\rangle.
    \end{equation}
    Finally, observe that replacing the string $\underline{\alpha}$ with identities should yield $g_{\id,\id,\dots}=1$, by definition.
    From this we can find the well known result $\lim_{l\to\infty}D\approx\lim_{l\to\infty}\lambda_{\rm max}^l=\varphi^l$, which can be used to obtain the final form for the expression for $t_{\underline{\alpha}}$
    \begin{equation}
        \label{seq:geq1}
        t_{\underline{\alpha}}=\lim_{l\to\infty}\frac{1}{\lambda_{\rm max}^{l_{\rm centre}}}\langle e_{\rm max}|\prod_{i\in{\rm center}}\widetilde{T}_{\alpha_i}|e_{\rm max}\rangle.
    \end{equation}
    
    In case of the PXP model we have, for the largest eigenvalue of the transfer matrix $\widetilde{T}_0$, $\lambda_{\rm max}=(1+\sqrt{5})/2=\varphi$, which is the golden ratio.
    This is directly connected to the scaling of the dimensionality of the constrained subspace with system size which scales as $\lim_{l\to\infty}D=\lim_{l\to\infty}\lambda_{\rm max}^l\sim\varphi^l$.
    We find the dominant eigenvector to take the form $|e_{\rm max}\rangle=|\varphi/\sqrt{2+\varphi},1/\sqrt{2+\varphi}\rangle$. 
    For example, for the trace of $\id_{2i}\otimes\sigma^z_{2i+1}\otimes\id_{2i+2}\otimes\sigma^z_{2i+3}$ operator we get:
    \begin{equation}
        g_{0,z,0,z}=\frac{\langle e_{\rm max}|\widetilde{T}_0\widetilde{T}_z\widetilde{T}_0\widetilde{T}_z|e_{\rm max}\rangle}{\lambda_{\rm max}^4}=3-\frac{6}{\sqrt{5}}.
    \end{equation}
    
    For the calculation of the cost function, it is sufficient to consider 10 spins for the \emph{center} region in Eq.~\eqref{seq:geq1}.
    Strictly speaking one could also get away with 8 by making use of translational invariance, since at worst the operator $G_{i,j}(t)$ that contributes to $S$ extends across 4 unit cells. 
    
    We now have all the necessary components to compute the relevant contributions of the cost function $S$ (recall we neglect terms that do not depend on $c_\eta$, since they are irrelevant to the calculation of the optimal values of $c_\eta$). 
    The $S$ obtained in this way is a function of the parent Hamiltonian parameters $a_1$, $a_2$, $b_1$, $b_2$ and their derivatives with respect to time, as well as driving Hamiltonian parameters $c_1$ and $c_2$. 
    One must then find the minimum of $S$ with respect to $c_1$ and $c_2$.
    As we already mentioned $S$ is a quadratic polynomial in $c_\eta$, so this is trivial to obtain.
    The resulting solutions can be found in~\footnote{See Supplemental Material for explicit expressions of the driving parameters $c_1$ and $c_2$.}.
    
    \subsection{Ground state minimization}
    \label{app:GS}
    
    In this section we will briefly discuss the derivation of the ground state CD approach from the full deviation form suggested by \textcite{sels2017}.
    We then discuss when this can reduce to a simple one-point function and finally describe the calculation for our example where we know the ground state in terms of an MPS representation. 
    Finally, we show some details on the calculation in the PXP model. 
    
    \subsubsection{Simplifying the cost function in the general case}
    \label{app:GS-derivation}
    \label{app:GS-onepoint}

    We begin with the expression for the cost function in the zero-temperature CD approach
    \begin{equation}
        S=\langle\psi(t)|G(t)^2|\psi(t)\rangle-\langle\psi(t)|G(t)|\psi(t)\rangle^2.
    \end{equation}
    In this section we show that the second term always vanishes.
    
    Expanding the original expression
    \begin{equation}
        \begin{split}
            &\hspace{2.5cm}\langle\psi(t)|G(t)|\psi(t)\rangle=\\
            &=\langle\psi(t)|\partial_tH_\text{p}(t)+iA(t)H_\text{p}(t)-iH_\text{p}(t)A(t)|\psi(t)\rangle,
        \end{split}
    \end{equation}
    it is immediately clear that the last 2 terms in the expectation value vanish due to choice of energy to be zero $H_\text{p}(t)|\psi(t)\rangle=0$ and $\langle\psi(t)| H_\text{p}(t)=0$. 
    This leaves the first term, that can be simplified using the relation
    \begin{equation}
        \label{seq:dh0}
        \partial_t(H_\text{p}(t)|\psi(t)\rangle)=(\partial_tH_\text{p}(t))|\psi(t)\rangle+H_\text{p}(t)(\partial_t|\psi(t)\rangle)=0,
    \end{equation}
    from which one obtains the following relation $(\partial_tH_\text{p}(t))|\psi(t)\rangle=-H_\text{p}(t)(\partial_t|\psi(t)\rangle)$.
    Substituting this into our expression and again accounting for $H_\text{p}(t)|\psi(t)\rangle=0$ we obtain our result
    \begin{equation}
        \langle\psi(t)|G(t)|\psi(t)\rangle=0.
    \end{equation}
    Thus we need to compute only the first term in the cost function
    \begin{equation}\label{Eq:action-simple}
        S=\langle\psi(t)|G(t)^2|\psi(t)\rangle.
    \end{equation}
    Note, that we always can choose $H_\text{p}(t)|\psi(t)\rangle=0$ without loss of generality. 
    
    We proceed to further simplify the expression for the zero temperature CD cost function by using the local structure of the parent Hamiltonian and driving operator $A$. 
    We note that this simplification omits the part of the expectation value that does not depend on $c_\eta$, similarly to what we did for trace CD. 
    To simplify the cost function, one has to explicitly write the terms entering the product of operators $G_i(t) G_j(t)$, and simplify them individually using the property $H_\text{p}(t)|\psi(t)\rangle=0$ and the fact that different local density operators commute with each other if $|i-j|$ is sufficiently large.
    This procedure results in the following expression for the cost function, that is similar to Eq.~(\ref{seq:1}),
    \begin{equation}
        S=\sum_j\sum_{k=-l_+-l_-+2}^{l_++l_--2}\langle\psi(t)|G_j(t)G_{j+k}(t)|\psi(t)\rangle.
    \end{equation}
    The summation limit depends on $l_\pm$ that are defined after Eq.~(\ref{G-simplified}) above.
    Using translational invariance with a unit cell of size $l_U$, the cost function can be further written as
    \begin{equation}
        S=\frac{l}{l_U}\sum_{j=1}^{l_U}\sum_{k=-l_+-l_-+2}^{l_++l_--2}\langle\psi(t)|G_j(t)G_{j+k}(t)|\psi(t)\rangle.
    \end{equation}
    
    \subsubsection{Calculation for the PXP model}
    \label{app:GS-calc}
    
    In case of the PXP model with a 2-site unit cell used throughout this paper we can then write
    \begin{equation}
        S=\frac{l}{2}\langle\psi|\widetilde{O}|\psi\rangle,
    \end{equation}
    where we defined
    \begin{equation}
        \widetilde{O}=\sum_{j=1}^{2}\sum_{k=-2}^{2}G_j(t)G_{j+k}(t),
    \end{equation}
    as the operator whose expectation value we must evaluate. 
    Since we are working in the thermodynamic limit this means we must simply find the contraction
    \begin{equation}
        \label{Eq:contract}        
        \includegraphics[width=0.8\linewidth]{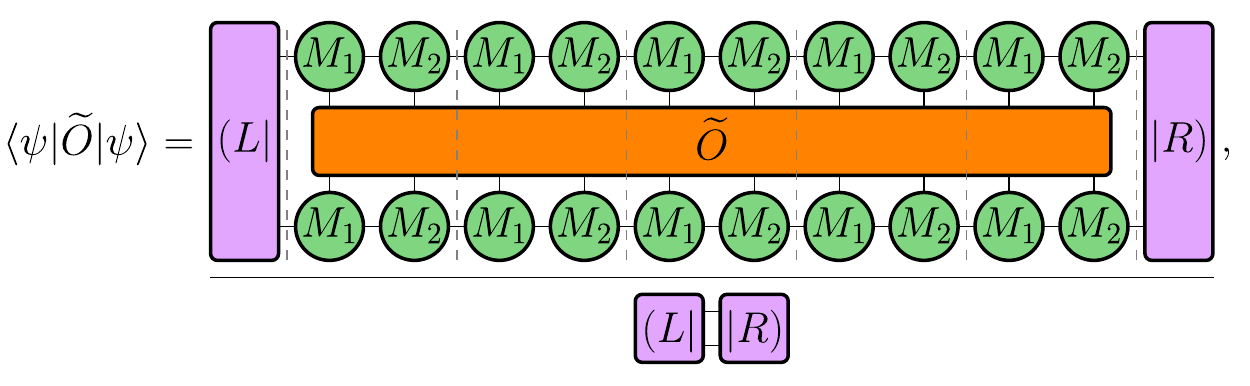}
    \end{equation}
    where $M_1$ and $M_2$ are the MPS tensors, $\widetilde{O}$ is the operator in question and $|L)$ and $|R)$ are the left and right dominant eigenvectors of the transfer matrix $\mathbb{T}$ respectively, which we define in the next paragraph.
    Note that here all vertical legs correspond to physical spin indices while the horizontal legs identify the auxiliary spaces -- in terms of the representation of the MPS in Eq.~\eqref{eq:mps} the physical dimensions are the ones corresponding to $|\hspace{-1mm}\uparrow\rangle$ and $|\hspace{-1mm}\downarrow\rangle$ and the row/column indices of the matrix correspond to the auxiliary dimensions of the MPS. 
    
    Due to translational invariance one can perform the calculation on only 4 unit cells as opposed to the 5 shown here by translating some terms by one unit cell, which may in some cases make the computation faster. 
    Similarly to the  case of trace CD calculation, this expression can be used in the thermodynamic limit where only dominant eigenvectors contribute. 
    Here we have defined the transfer matrix to move us by one unit cell
    \begin{equation}
        \label{seq:T-GS}
        \mathbb{T}=\prod_{i=1}^2\left(M_i^{\uparrow}\otimes \overline{M_i^\uparrow}+M_i^{\downarrow}\otimes\overline{M_i^\downarrow}\right),
    \end{equation}
    where $M_i^{\uparrow/\downarrow}$ are the corresponding matrices of tensor $M_i$ at the respective physical indices.
    Note also that this takes the form of a matrix once the two auxiliary indices on each side are merged.
    For the transfer matrix we obtain
    \begin{equation}
        \mathbb{T}=\begin{pmatrix}
            \cos^2\theta_1\cos^2\theta_2+\sin^2\theta_1 & 0 & 0 & \cos^2\theta_1\sin^2\theta_2\\
            \cos\theta_1\cos^2\theta_2 & 0 & 0 & \cos\theta_1\sin^2\theta_2\\
            \cos\theta_1\cos^2\theta_2 & 0 & 0 & \cos\theta_1\sin^2\theta_2\\
            \cos^2\theta_2 & 0 & 0 & \sin^2\theta_2
        \end{pmatrix}.
    \end{equation}
    Eigenvectors $|L)$ and $|R)$ are then the left and right eigenvectors with eigenvalue 1, $(L|\mathbb{T}=(L|$ and $\mathbb{T}|R)=|R)$. Their explicit form reads:
    \begin{align}
        \label{seq:L-GS}
        |L)&=\frac{1}{\sqrt{\sin^4\theta_1\sin^4\theta_2+\cos^4\theta_2}}\begin{pmatrix}\cos^2\theta_2 \\ 0 \\ 0 \\ \sin^2\theta_1\sin^2\theta_2\end{pmatrix},
        \\
        \label{seq:R-GS}
        |R)&=\frac{1}{\sqrt{3+\cos2\theta_1}}\begin{pmatrix}1\\\cos\theta_1\\\cos\theta_1\\1\end{pmatrix}.
    \end{align}
    Note that there exists a single point $\theta_1=\pi/2+k\pi$ and $\theta_2=\pi/2+k\pi$ when the dominant eigenvalue becomes degenerate, however since that point is disconnected from the rest of the MPS manifold, we do not discuss it further. 
    
    Using this one must simply perform the contraction depicted in Eq.~(\ref{Eq:contract}) and minimize the result with respect to the driving parameters $c_\eta$ in order to obtain the driving parameters from the ground state CD approach.
    The resulting expressions for the driving parameters can be found in~\cite{Note3}. 
    
    \vspace{-1mm}
    \section{Calculation with the leakage approach in the PXP model}
    \label{app:leak-calc}
    
    A natural extension of the ground state CD approach is leakage minimization, which is closely related. 
    In this approach we must minimize the cost function defined as $S=\delta^2$ from Eq.~(\ref{eq:leakage}).
    Here the leakage $\delta$ represents the norm of the vector pointing away from our MPS trajectory. 
    Unlike what we observed in the ground state CD approach, this does not reduce to a one-point function. 
    Nevertheless the expression can still be computed exactly. 
    
    Expanding the equation for the leakage we get
    \begin{multline}
        \label{seq:lmin-1}
            S=\frac{1}{l}\Big[{\rm Re}\langle\partial_t\psi(t)|\partial_t\psi(t)\rangle_c
            -2{\rm Im}\langle\partial_t\psi(t)|A(t)|\psi(t)\rangle_c
            \\+\langle\psi(t)|A(t)^2|\psi(t)\rangle_c\Big],
    \end{multline}
    where we introduced a shorthand notation $|\partial_t\psi(t)\rangle=\partial_t|\psi(t)\rangle$. 
    Note that the use of connected correlations in the above expression is the result of requiring the metric to be invariant under multiplication by a global phase, which we discuss in a bit more detail in the next section. 
    As we will see this will also exactly cancel all terms proportional to $l^2$, leaving only terms proportional to $l$ in the correlations themselves. 
    
    To reduce the number of terms in the calculation we introduce 2-site tensors $(M_{12})_{a_l,s_1,s_2,a_r}=\sum_{a_m}(M_1)_{a_l,s_1,a_m}(M_2)_{a_m,s_2,a_r}$, where $a_m$ traces out the auxiliary space between $M_1$ and $M_2$. 
    The remaining indices are $s_1$ and $s_2$ for the physical degree of freedom on site 1/2 respectively, $a_l$ for the auxiliary link to the left of site 1 and $a_r$ for the auxiliary link to the right of site 2. 
    In this formulation our problem has a unit cell size of 1, greatly simplifying all following expressions. 
    Let us first look at the first term ${\rm Re}\left(\langle\partial_t\psi(t)|\partial_t\psi(t)\rangle\right)$. 
    Using the chain rule for the derivatives this can be expressed using MPS as
    \begin{equation}
        \includegraphics[width=1.0\linewidth]{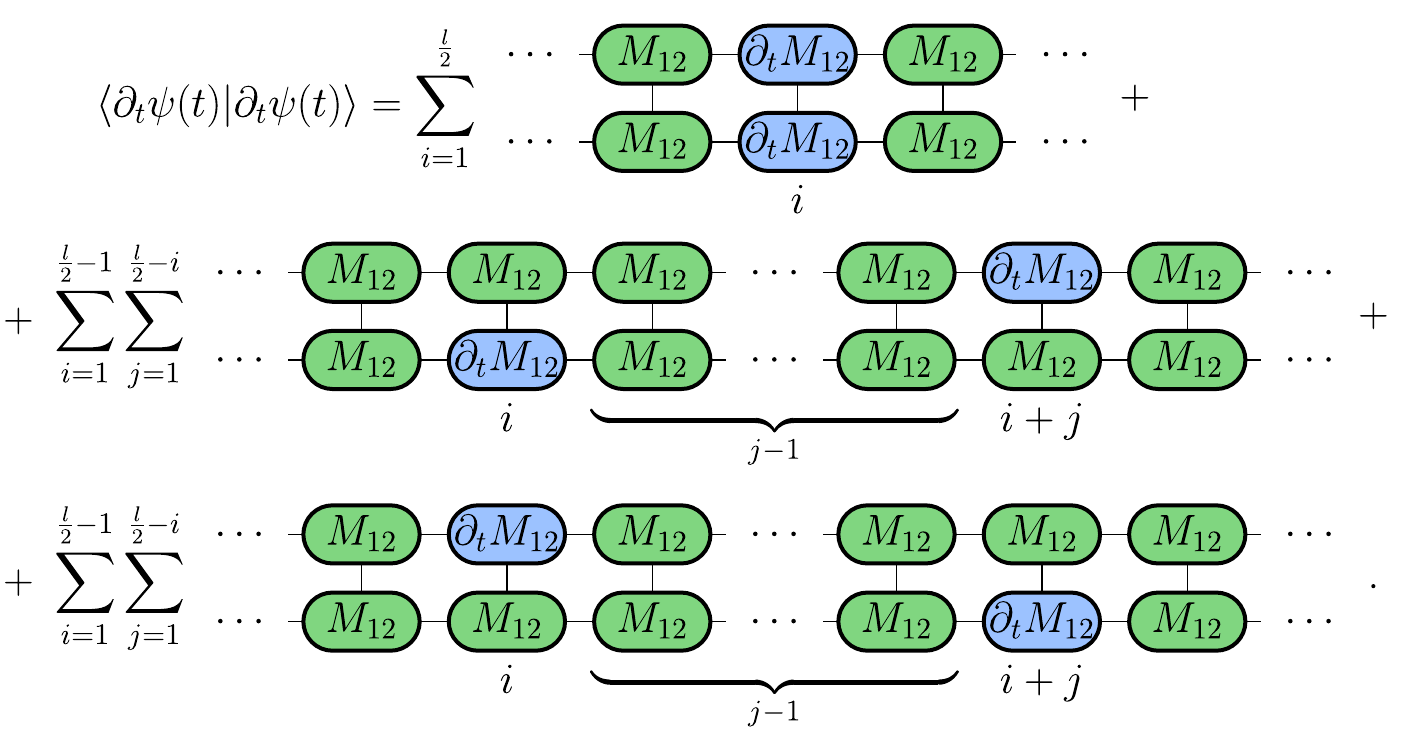}
    \end{equation}
    Here we introduced the derivative terms as 
    \begin{equation}
        \begin{split}
            (\partial_tM_{12})_{a_l,s_1,s_2,a_r}=&\sum_{a_m}[(\partial_tM_1)_{a_l,s_1,a_m}(M_2)_{a_m,s_2,a_r}+\\&+(M_1)_{a_l,s_1,a_m}(\partial_tM_2)_{a_m,s_2,a_r}].
        \end{split}
    \end{equation}
    We can already see that the sum in $j$ resembles a geometric series in the transfer matrix $\mathbb{T}$ defined in Eq.~\eqref{seq:T-GS}.
    Since the transfer matrix has an eigenvalue equal to 1, this series is divergent. 
    We can circumvent this issue by projecting out the divergent part and treating it separately
    \begin{equation}
        \lim_{l\to\infty}\sum_{n=0}^{l}\mathbb{T}^n=\lim_{l\to\infty}l\mathcal{P}+\underbrace{\mathcal{Q}(\id-\mathcal{Q}\mathbb{T}\mathcal{Q})^{-1}\mathcal{Q}}_{\mathcal{T}^{-1}},
    \end{equation}
    where we define the projectors to the dominant sub-space $\mathcal{P}=\frac{|R)(L|}{(R|L)}$ and it's complement $\mathcal{Q}=\id-\mathcal{P}$.
    Here $|L)$ and $|R)$ are the left/right dominant eigenvectors of our transfer matrix defined in Eqs.~(\ref{seq:L-GS})-(\ref{seq:R-GS}). 
    Using this it is clear we must compute the following expression, represented in terms of tensors as 
    \begin{widetext}
        \begin{equation}
            \includegraphics[width=.94\linewidth]{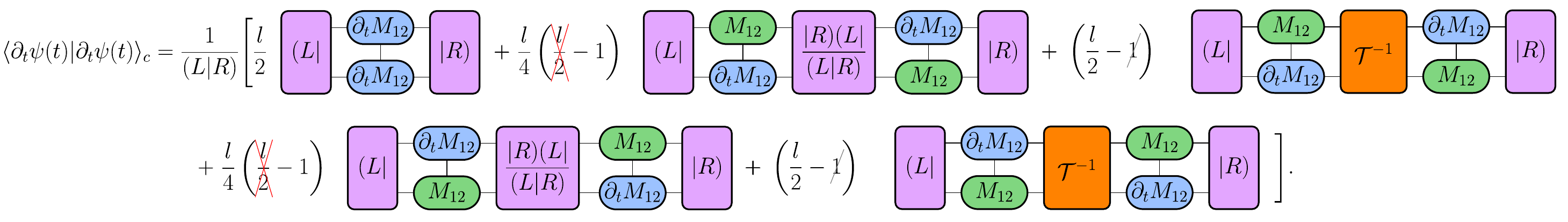}
        \end{equation}
    \end{widetext}
    Here the terms crossed out in red cancel out with the disconnected correlators and the the prefactor $(\frac{l}{2}-1)$ can be replaced by $l/2$ in the thermodynamic limit. 
    To express this in a more concise form let us define several matrices
    \begin{equation}
        \includegraphics[width=0.9\linewidth]{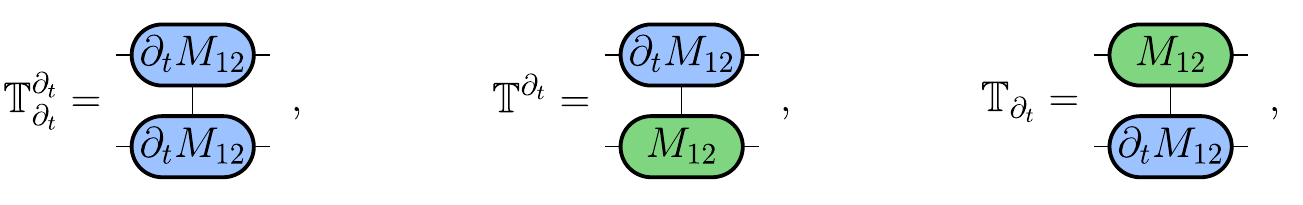}
    \end{equation}
    where all left-facing and right-facing legs are merged into a single row or column index respectively. 
    Using this we can write the expression as
    \begin{equation}
        \begin{split}
            \langle\partial_t\psi(t)|\partial_t\psi(t)\rangle_c=-\frac{l}{2}\frac{(L|\mathbb{T}_{\partial_t}|R)(L|\mathbb{T}^{\partial_t}|R)}{(L|R)^2}&+\\
            +\frac{l}{2}\frac{(L|\mathbb{T}_{\partial_t}^{\partial_t}+\mathbb{T}_{\partial_t}\mathcal{T}^{-1}\mathbb{T}^{\partial_t}+\mathbb{T}^{\partial_t}\mathcal{T}^{-1}\mathbb{T}_{\partial_t}|R)}{(L|R)}&.
        \end{split}
    \end{equation}
    
    We can now apply the same approach to the other connected correlation functions in the expression for the leakage. Defining the  density of the driving Hamiltonian $A$ with respect to the unit cell, $\widetilde{A}$, which, after merging the auxiliary indices, can be written as
    \begin{equation}
        \widetilde{A}=c_1P\otimes\sigma^x\otimes P\otimes\id+c_2\id\otimes P\otimes\sigma^x\otimes P,
    \end{equation}
    we introduce  several auxiliary matrices (again merging the left/right auxiliary indices as was done previously)
    \begin{equation}
        \includegraphics[width=.9\linewidth]{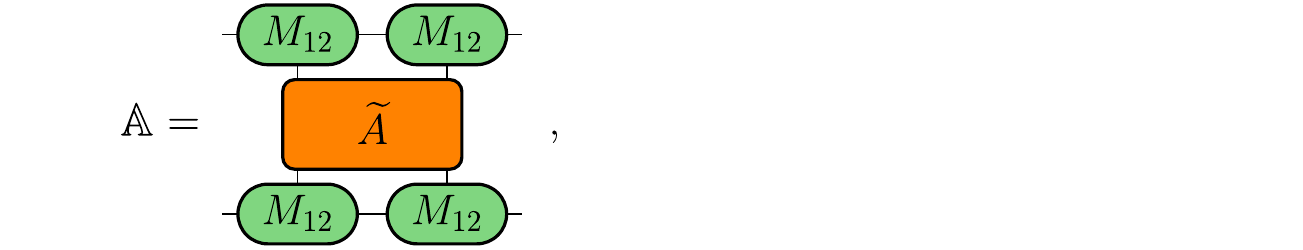}
    \end{equation}
    \begin{equation}
        \includegraphics[width=.9\linewidth]{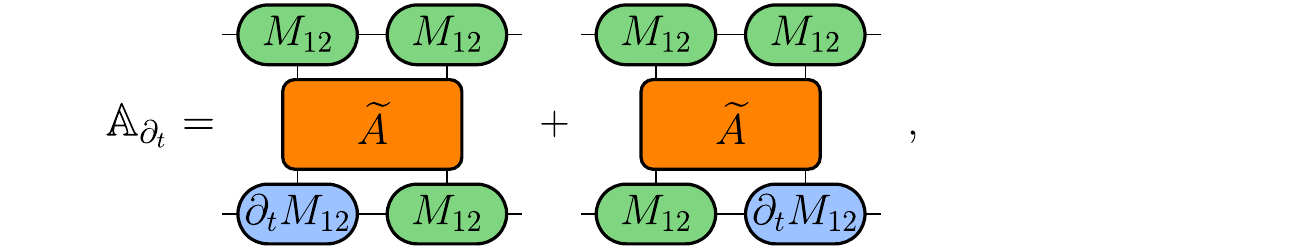}
    \end{equation}
    \begin{equation}
        \includegraphics[width=.9\linewidth]{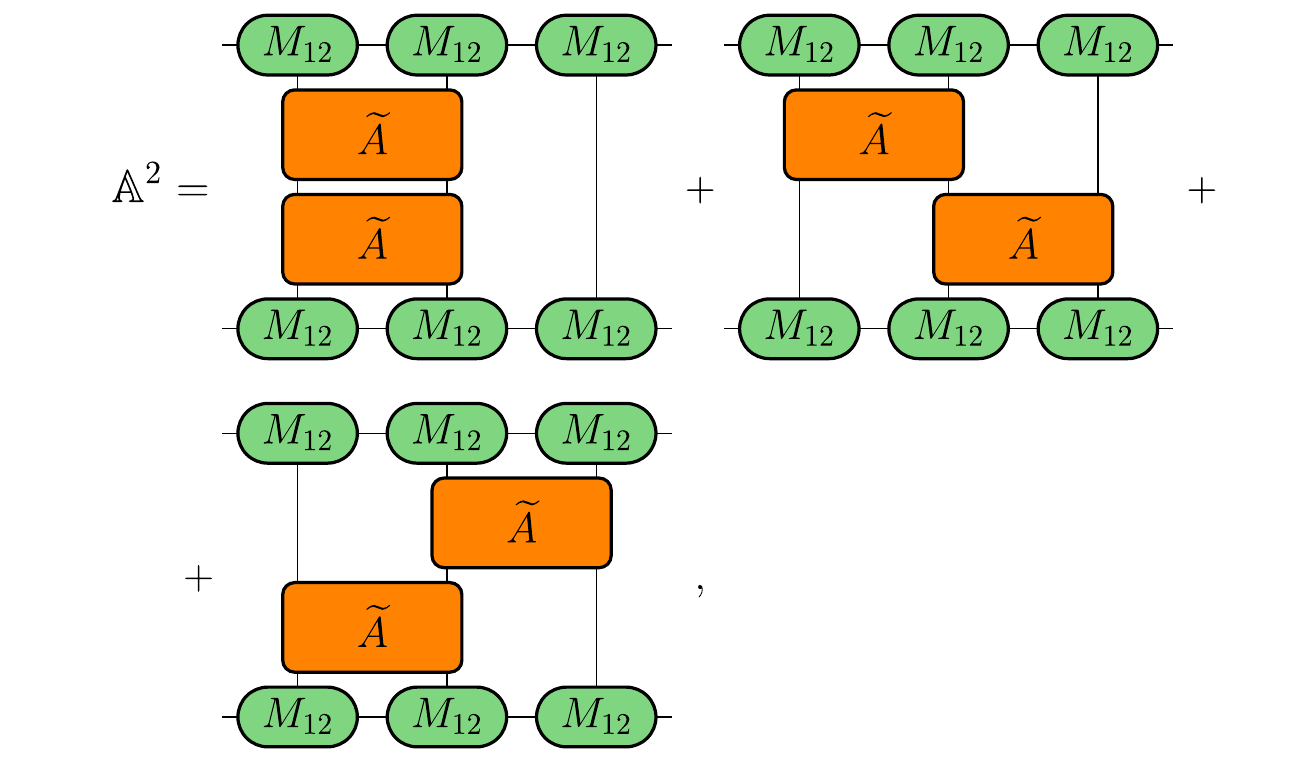}
    \end{equation}
      Using this one may then obtain the expressions for the other two connected correlators in the thermodynamic limit
    \begin{eqnarray}
        \nonumber
            \langle\partial_t\psi(t)|A(t)|\psi(t)\rangle_c=-2\frac{l}{2}\frac{(L|\mathbb{A}|R)(L|\mathbb{T}_{\partial_t}|R)}{(L|R)^2}&+\\\label{seq:t1}
            \frac{l}{2}\frac{(L|\mathbb{A}_{\partial_t}+\mathbb{T}_{\partial_t}\mathcal{T}^{-1}\mathbb{A}+\mathbb{A}\mathcal{T}^{-1}\mathbb{T}_{\partial_t}|R)}{(L|R)},&
            \\
        \nonumber
            \langle\psi(t)|A(t)^2|\psi(t)\rangle_c=-3\frac{l}{2}\frac{(L|\mathbb{A}|R)^2}{(L|R)^2}&+\\\label{seq:t2}
            \frac{l}{2}\frac{(L|\mathbb{A}^2+2\mathbb{A}\mathcal{T}^{-1}\mathbb{A}|R)}{(L|R)}.&
    \end{eqnarray}
    The factors in the first term, 2 and 3 respectively, represent the number of unit cells where the two operators, $\partial_tM_{12}$ and $\widetilde{A}$ in Eq.~\eqref{seq:t1}, and $\widetilde{A}$ and $\widetilde{A}$ in Eq.~\eqref{seq:t2}, would overlap and must be subtracted. 
       
    In our example all these correlation functions can be obtained analytically and then combined to find the analytical expression for the cost function $S=\delta^2$
         \begin{widetext}
        \begin{equation}
            \begin{split}
                S=&\Big[
                c_2(t)^2\left(7+8\cos2\theta_1(t)+2\cos4\theta_1(t)\cos^2\theta_2(t)-\cos2\theta_2(t)\right)+\\
                &~+c_1(t)^2\left(7+8\cos2\theta_2(t)+2\cos4\theta_2(t)\cos^2\theta_1(t)-\cos2\theta_1(t)\right)+64c_1(t)c_2(t)\cos^2\theta_1\cos^2\theta_2\sin\theta_1\sin\theta_2+\\
                &~+32c_2(t)\cos^2\theta_1(t)(\theta'_2(t)\cos\theta_1(t)+\theta'_1(t)\cos\theta_2(t)\sin\theta_1(t)\sin\theta_2(t))+\\
                &~+32c_1(t)\cos^2\theta_2(t)(\theta'_1(t)\cos\theta_2(t)+\theta'_2(t)\cos\theta_1(t)\sin\theta_2(t)\sin\theta_1(t))+\\
                &~+8\left(\theta'_1(t)^2(1+\cos2\theta_2)+\theta'_2(t)^2(1+\cos2\theta_1)\right)\Big]/\Big[8(3+\cos2\theta_1(t)+2\cos2\theta_2(t)\sin^2\theta_1(t))\Big].
            \end{split}
        \end{equation}
        From this we can obtain the solution for the driving parameters $c_1(t)$ and $c_2(t)$ for an arbitrary trajectory in the MPS manifold, defined by $(\theta_1(t),\theta_2(t))$ by minimizing the above cost function with respect to the driving parameters 
        \begin{equation}
            \begin{split}
                c_{1,2}(t)=&2\Big[-2\theta'_{1,2}(t)(9+6\cos2\theta_{1,2}(t)+\cos4\theta_{1,2}(t))\cos\theta_{2,1}(t)-8\theta'_{1,2}(t)\cos2\theta_{1,2}(t)\cos3\theta_{2,1}(t)\sin^2\theta_{1,2}(t)+\\&+\theta'_{2,1}(t)(2\sin2\theta_{1,2}(t)+7\sin4\theta_{1,2}(t))\sin\theta_{2,1}(t)+8\theta'_{2,1}(t)\cos\theta_{1,2}(t)\sin^3\theta_{1,2}(t)\sin3\theta_{2,1}(t)\Big]/\\&/\Big[29+3\cos4\theta_{1,2}(t)+32\cos2\theta_{1,2}(t)\cos2\theta_{2,1}(t)+6\cos4\theta_{2,1}(t)\sin^22\theta_{1,2}(t)\Big].
            \end{split}
        \end{equation}
    \end{widetext}
    
    Although the cost function can be computed analytically in the present example, such calculation will very quickly become infeasible upon increasing the bond dimension. 
    The particularly problematic parts are the finding the pseudo-inverse of the transfer matrix $\mathcal{T}^{-1}$ and the left/right eigenvectors $(L|$ and $|R)$ of the transfer matrix $\mathbb{T}$. 
    Importantly, even when analytic calculation is infeasible, one can perform our procedure at any point in the phase space numerically.
    As a result, this approach can easily be applied to MPS states with large bond dimension and/or unit cells. 
    
    \section{Relation between leakage minimization and ground state CD} 
    \label{app:leak-vs-GS}
    
    Recall that the ground state CD approach amounts to minimizing the cost function defined in Eq.~\eqref{Eq:action-simple}.
    Since $G(t)$ is Hermitian, this can then be rewritten as
    \begin{equation}
        S=\left|\left(\partial_tH_\text{p}(t)+i[A(t),H_\text{p}(t)]\right)|\psi(t)\rangle\right|^2.
    \end{equation}
    Expanding the commutator we find that one of the contributions vanishes $iA(t)H_\text{p}(t)|\psi(t)\rangle=0$, due to $H_\text{p}(t)|\psi(t)\rangle=0$. 
    We are then left with
    \begin{equation}
        S=\left|\left(\partial_tH_\text{p}(t)-iH_\text{p}(t)A(t)\right)|\psi(t)\rangle\right|^2.
    \end{equation}
    Finally, recall from Eq.~\eqref{seq:dh0} that we have $\partial_tH_\text{p}(t)|\psi(t)\rangle=-H_\text{p}(t)\partial_t|\psi(t)\rangle$.
    Using this we rewrite the ground state CD cost function as
    \begin{equation}
        S=\left|H_\text{p}(t)\left[\partial_t|\psi(t)\rangle+iA(t)|\psi(t)\rangle\right]\right|^2,
    \end{equation}
    where the minus is absorbed by the absolute value. 
    Comparing this with the expression for the leakage, $\delta^2$
    $$
        \delta^2=\frac1l\left|\partial_t|\psi(t)\rangle+iA|\psi(t)\rangle\right|^2.
    $$
    Is it immediately clear that the ground state CD cost function is nothing but the leakage weighted by the parent Hamiltonian (the missing $1/l$ factor is not relevant to the optimization). 
    
    \section{TDVP, leakage and rescaled leakage}
    \label{app:leak-vs-tdvp}
    
    \begin{figure*}[ht!]
        \centering
        \label{sfig:triangles}
        \includegraphics[height=4.5cm]{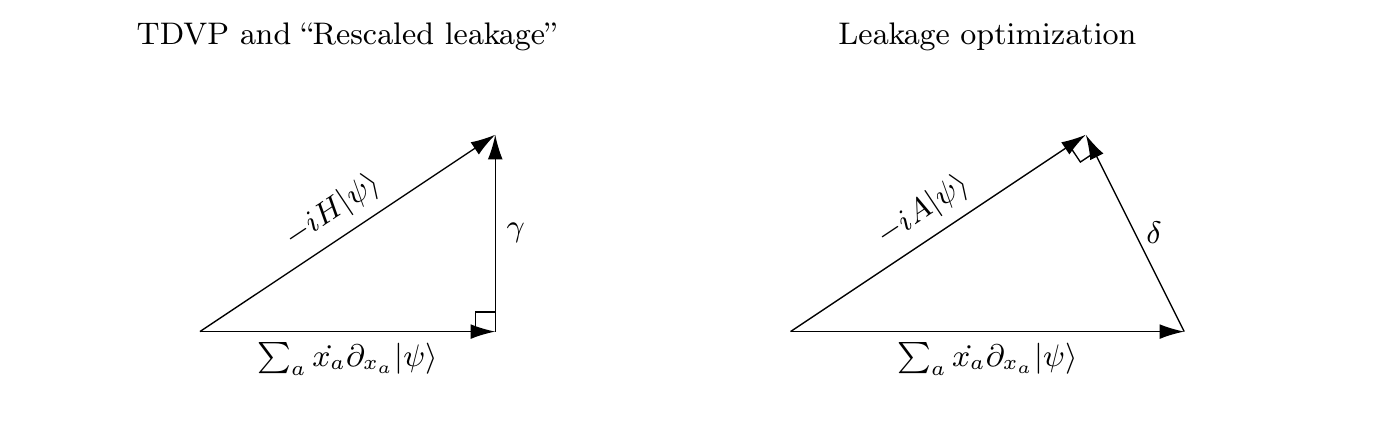}
        \caption{
            Schematic representation of the difference between TDVP and leakage minimization.
            Due to the similar nature of optimization in both cases one gets a right angle opposite to the side that is fixed, the difference is only in the side that is fixed.
            In case of TDVP we optimize $\sum_a\dot{x_a}\partial_{x_a}|\psi\rangle$ while keeping $-iA|\psi\rangle$ fixed whereas in our leakage minimization approach the two swap roles. 
            This suggests that there is an alternative formulation (dubbed ``rescaled leakage'' in what follows), where instead of minimizing leakage we choose $c_i$ such that the projection of $iA'|\psi\rangle$ to $\partial_t|\psi\rangle$ would have the exact same norm as $\partial_t|\psi\rangle$. 
            In this way the dynamics in the trajectory manifold would be exact, when possible. 
        }
    \end{figure*}
    
    Both TDVP and our leakage based optimization rely on the minimization of leakage and an MPS formulation. 
    Let us define the MPS as $|\psi(t)\rangle=|\psi(\{x_j(t)\})\rangle$, with $x_j$ being the real parameters of the MPS ansatz. 
    In the TDVP approach, we take the Hamiltonian $H$ that generates the dynamics and find the vector in the tangent space of the MPS manifold, spanned by the vectors $\partial_{x_j}|\psi(t)\rangle$, which gives rise to the lowest leakage.
    This is done by simply projecting $-iA|\psi(t)\rangle$ to the tangent space and requiring that the projection is described exactly
    \begin{equation}
        {\rm Re}\langle\partial_{x_j}\psi(t)|\partial_t\psi(t)\rangle_c+{\rm Re}\left[i\langle\partial_{x_j}\psi(t)|A(t)|\psi(t)\rangle_c\right]=0;~\forall j.
    \end{equation}
    Substituting $|\partial_t\psi(t)\rangle=\sum_j\dot{x}_j\partial_{x_j}|\psi(t)\rangle$ one can invert the system of equations to obtain expressions for $\dot{x}_j$. 
    Similarly our leakage approach simply inverts the roles, we now minimize the distance between the desired direction in the MPS manifold $|\partial_t\psi(t)\rangle$ and the space spanned by $-iA_j|\psi(t)\rangle$.
    While we formulate this as a direct minimization of leakage
    \begin{equation}
        \min_{c_\eta}\left[\frac{1}{l} {\left|\partial_t|\psi(t)\rangle+iA(t)|\psi(t)\rangle\right|^2}\right],    \end{equation}
    it is easy to see that the minimization of this expression is equivalent to
    \begin{equation}
        \label{seq:lmin-proj}
        {\rm Re}\left[i\langle\psi(t)|A_j|\partial_t\psi(t)\rangle_c\right]-\langle\psi(t)|A_jA(t)|\psi(t)\rangle_c=0~;~\forall j.
    \end{equation}
    This of course is nothing but the projection of $|\partial_t\psi(t)\rangle$ to the manifold spanned by $-iA_j|\psi(t)\rangle$, where we fix $c_\eta$ such that the difference between the projection and our solution in the manifold is zero. 
    This suggests a simple geometric description presented in Fig.~5.
    
    This gives rise to an interesting question; is it possible to choose $c_\eta$ such that the projection to the trajectory manifold would be exact (see left schematic in Fig.~5 with $H\equiv A$) and thus the dynamics within that manifold would be exact at the expense of higher overall leakage.
    Here the trajectory manifold corresponds to the 1-dimensional sub-manifold of states defined by $|\psi(t)\rangle$ within the full MPS manifold. 
    
    In order to explore this idea let us first briefly discuss the various spaces involved here. 
    Firstly, let us define the Hilbert space spanning all states $\mathcal{H}$ with the inner product $\langle\alpha,\beta\rangle=\langle\beta|\alpha\rangle$. 
    This of course is nothing but $\mathbb{C}^n$ for finite systems, but naturally remains a Hilbert space when $n$ tends to infinity.
    Here $n$ is the dimensionality of our system -- for a spin-1/2 chain of length $l$ this is simply $n=2^l$.
    Note however that this space is not ideal for our approaches, as has been noted in the original time dependent variational principle works and in the geometric formulation of quantum mechanics. 
    The issue arises with the length of $|\psi\rangle-e^{-i\varphi}|\psi\rangle$, physically these are the same state and so ideally the distance between them should be zero. 
    As mentioned, this has been studied in the past in terms of the geometric formulation of quantum mechanics~\cite{Fubini1904,Study1905,Kibble1979}.
    Our quantum states ``live'' in the complex projected space $CP^{n-1}$ (note that for instance $CP^1$ is just the Bloch sphere).
    Indeed this space is simply the quotient of a unit sphere under the action of $U(1)$, such that $CP^{n-1}=S^{2n}/U(1)$, the sphere coming from normalization and the $U(1)$ from invariance of states with respect to a global phase.
    
    In the complex projected space $CP^{n-1}$ we can define the Fubini-Study metric~\cite{Fubini1904,Study1905} 
    \begin{equation}
        \gamma(\psi,\phi)=\arccos\sqrt{\frac{\langle\psi|\phi\rangle\langle\phi|\psi\rangle}{\langle\phi|\phi\rangle\langle\psi|\psi\rangle}},
    \end{equation}
    where the normalization terms can of course be omitted. 
    One may then assign a tangent space $T_{|\psi\rangle}CP^{n-1}$ to each point $|\psi\rangle$ of the complex projected space with a metric tensor
    \begin{equation}
        g_{|\psi\rangle}(|\alpha\rangle,|\beta\rangle)={\rm Re}\left[\langle\alpha|\beta\rangle-\langle\alpha|\psi\rangle\langle\psi|\beta\rangle\right].
    \end{equation}
    Here $|\alpha\rangle$ and $|\beta\rangle$ are vectors in the tangent space $T_{|\psi\rangle}CP^{n-1}$. 
    Clearly this metric tensor is nothing but a connected correlation function, so we will generally use the notation $g_{|\psi\rangle}(|\alpha\rangle,|\beta\rangle)={\rm Re}\langle\beta|\alpha\rangle_c$. 
    Importantly, the metric tensor gives rise to an inner product, which in turn allows for the definition of angles and distances in the tangent space.
    Additionally, since the tangent space $T_{|\psi\rangle}CP^{n-1}$ is complete, it follows that it is a Hilbert space. 
    This property will prove useful since it allows us to use simple geometric arguments to obtain $c_\eta$ such that the dynamics within the trajectory manifold are accurate.
    
    Indeed, since both the tangent space of the MPS manifold ($T_{|\psi\rangle}\Psi$) spanned by $\partial_{x_j}|\psi\rangle$ and the driving manifold $\mathcal{A}$ spanned by $-iA_j|\psi\rangle$ are sub-spaces of $T_{|\psi\rangle}CP^{n-1}$, we can make some simple observations.
    Firstly, we can decompose any vector in this space into a length $r$ and direction $\vec{e}$ components. 
    From this, since both sub-spaces are vector spaces themselves, it is clear that for any direction $\vec{e}_{T_{|\psi\rangle}\Psi}$ in the tangent space of the MPS manifold there is a corresponding direction $\vec{e}_{\mathcal{A}}$ in the driving subspace, which is nothing but the projection of $\vec{e}_{T_{|\psi\rangle}\Psi}$ onto $\mathcal{A}$. 
    This means that in order to obtain the driving parameters with correct projection on the trajectory manifold, we can simply re-scale the solution from the leakage minimization approach.
    Taking the solution for leakage minimization, it follows from Eq.~\eqref{seq:lmin-proj}
    \begin{equation}
        \langle\psi(t)|A(t)A(t)|\psi(t)\rangle_c=-{\rm Im}\langle\psi(t)|A(t)|\partial_t\psi(t)\rangle_c.
    \end{equation}
    Using this and basic calculus we obtain the rescaling factor $\omega$
    \begin{equation}
        \omega=\frac{\langle\partial_t\psi(t)|\partial_t\psi(t)\rangle_c}{\langle\psi(t)|A(t)A(t)|\psi(t)\rangle_c},
    \end{equation}
    which we can use to obtain the rescaled coefficients $c_{\eta,{\rm P}}=\omega c_\eta$. 
    In this expression $A(t)$ represents the solution from leakage minimization, which is used to obtain the rescaled driving Hamiltonian $A_P(t)$. 
    Note that cases where the two sub-spaces are orthogonal to one-another are not considered here, since there in this case optimization is not possible. 
    In the following, we will mark this approach as \emph{rescaled leakage}. 
    
    Naturally this is not equivalent to applying standard TDVP to the problem and then attempting to solve the minimization. 
    Such an approach would underestimate the leakage, since we would essentially only be minimizing the leakage within the tangent space of the MPS manifold. 
    Additionally, there may be dimensions in $\mathcal{A}$ orthogonal to $T_{|\psi\rangle}\Psi$, which would simply be projected out, leaving free parameters in the solutions. 
    Indeed we can easily create such a TDVP approach, which would give for the cost function
    \begin{equation}
        S=2{\rm Im}\langle A_{\mathcal{P}}(t)\overrightarrow{\partial_t}\rangle_c+\langle A_{\mathcal{P}}(t)A_{\mathcal{P}}(t)\rangle_c+{\rm constant},
    \end{equation}
    where $A_{\mathcal{P}}(t)=\mathcal{P}_{T_{|\psi\rangle}\Psi}A(t)\mathcal{P}_{T_{|\psi\rangle}\Psi}$ is nothing but $A(t)$ projected to the tangent space of the MPS manifold.
    Here the projector to the tangent space is defined as $\mathcal{P}_{T_{|\psi\rangle}\Psi}=\sum_{i,j}\left(G^{-1}\right)_{i,j}|\partial_{x_i}\psi(t)\rangle\langle\partial_{x_j}\psi(t)|$, with $G_{i,j}=\langle \partial_{x_i}\psi(t)|\partial_{x_j}\psi(t)\rangle$ the Gram matrix and $G^{-1}$ a pseudo-inverse.
    Note that $\mathcal{P}_{T_{|\psi\rangle}\Psi}|\psi(t)\rangle=|\psi(t)\rangle$ since $|\psi(t)\rangle$ is the point of the MPS manifold at which the tangent space is taken. 
    Similarly all derivatives $|\partial_{x_i}\psi(t)\rangle$ are also vectors in the tangent space and hence one also has  $\mathcal{P}_{T_{|\psi\rangle}\Psi}|\partial_{x_i}\psi(t)\rangle=|\partial_{x_i}\psi(t)\rangle$. 
    It is of course clear that this is not equivalent to the approach described before and one would expect it to perform worse on account of projecting the operator $A(t)$ to the tangent space. 
    
    \section{Additional trajectory deformations}
    \label{app:trajectories}
    
    \begin{figure*}[ht!]
        \centering
        \includegraphics[width=0.8\linewidth]{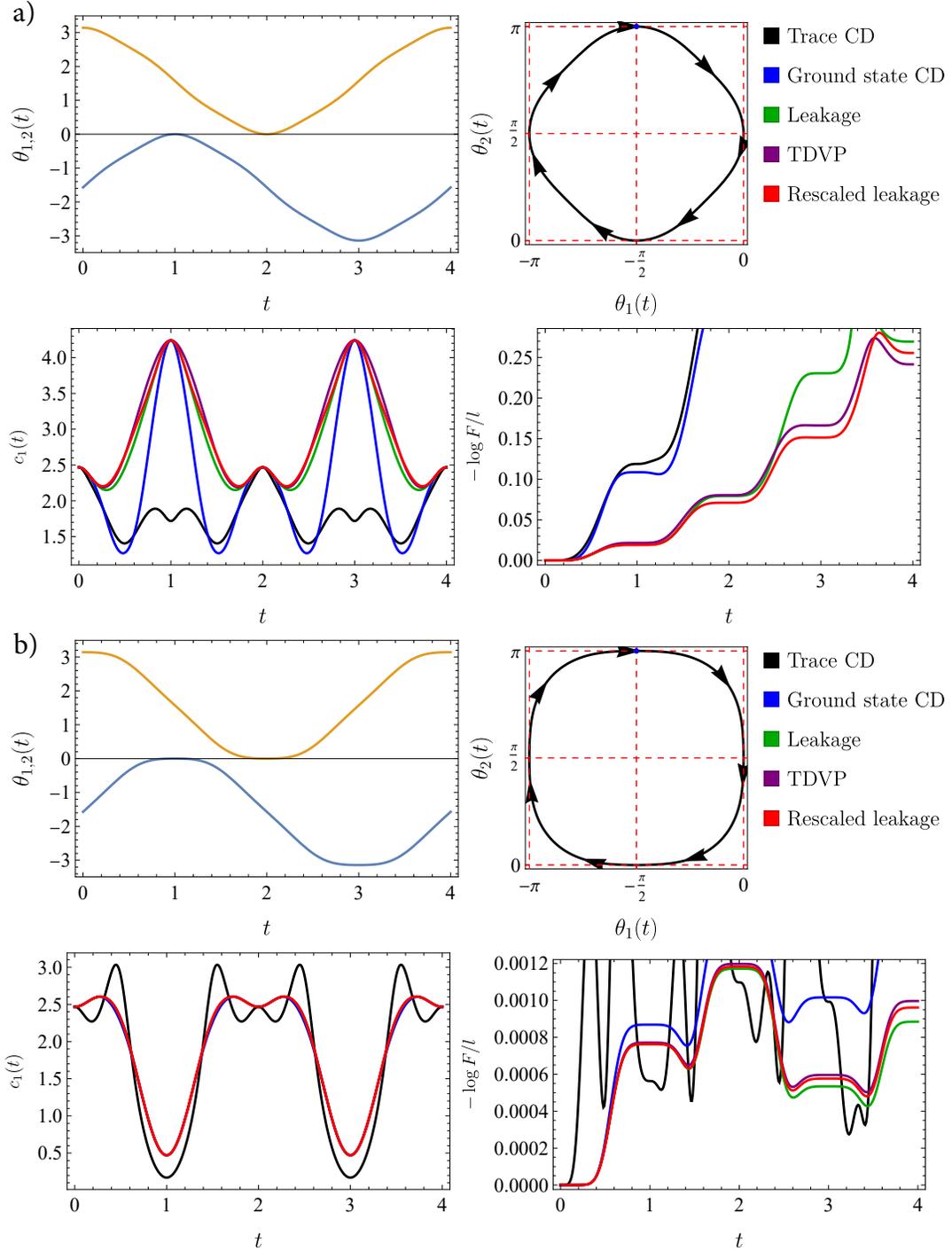}
        \caption{ \label{sfig:trajs1}
            We compare the performance of different approaches for the deformations of the circular trajectory marked in Fig.~\ref{fig:scar2} with $\varepsilon_1=+0.05$ in panel (a) and $\varepsilon_1=-0.05$ in panel (b), and $\varepsilon_2=0$ in both cases.
            In both examples, the leakage based approaches outperform ground state CD which in turn outperforms trace CD. 
            We observe that when leakage, and in turn $-\log F/l$, are small (see \textbf{b)}), leakage minimization is the better approach.
            However, when this is not the case as is shown in panel (a), the rescaled leakage approach appears to perform somewhat better. 
            All data was obtained with iTEBD using a maximal bond dimension of $\chi=512$ and a time step of $\delta t=10^{-5}$. 
        }
    \end{figure*}
    
    In this section we present several additional examples of trajectories described by Eqs.~(\ref{eq:angles})-(\ref{eq:deform}) in the main text. 
    To this end, Fig.~\ref{sfig:trajs1} compares different approaches for the trajectory with large leakage in panel (a) and for the trajectory with a small leakage in panel (b). 
    We emphasize that in panel (a) of Fig.~\ref{sfig:trajs1} fidelity is low at all times $t\gtrsim1$. 
    Indeed, to put things into perspective, at $-\log F/l=0.05$ the fidelity at $l=16$ is only $F\approx0.45$. 
    
    In both panels of Fig.~\ref{sfig:trajs1} we can observe that the leakage approach consistently outperforms trace and ground state CD approaches. 
    Among the two CD approaches, ground state CD has  better performance, which is in line with expectations.
    Surprisingly, we observe that the rescaled leakage approach, which we describe in Appendix~\ref{app:leak-vs-tdvp}, may outperform the leakage approach. 
    
    Since the rescaled leakage approach generally has higher instantaneous leakage, the fact that it performs better compared to leakage approach is surprising. 
    We note that this happens only once fidelity degrades sufficiently. 
    In other words, as long as the quantum wave function follows the desired trajectory with high fidelity, leakage minimization is the best approach. 
    However, if the wave function deviates sufficiently from the trajectory (even within the trajectory sub-manifold), the rescaled approach starts performing somewhat better.
    One possible explanation of such better performance is the better synchronization of the assumed dynamics within the trajectory manifold with those that actually take place in the quantum evolution with our driving parameters.
    Put differently, at least the projection of the state to the trajectory manifold is correct and hence our driving, which is calculated using that state, could be more trustworthy. 
    
    \section{Global trajectory optimization and Euler-Lagrange equations}
    \label{app:euler-lagrange}
    
    In this work we have mostly shown how to find the optimal driving parameters once a trajectory has been fixed, which was simply done by solving the system of equations
    \begin{equation}
        \partial_{c_\eta}S=0,
        \quad
        \forall\eta\in\{1,\dots,n\},
    \end{equation}
    where $n$ is the number of driving parameters (in our example this was 2). 
    If instead we wish to find the optimal trajectory from some point in the MPS manifold $|\psi(0)\rangle$ to some other point in the MPS manifold $|\psi(\tau)\rangle$, we would have to optimize
    \begin{equation}
        {\cal S}[c_\eta(t),x_k(t)]=\int_0^\tau {\rm d}t S,
    \end{equation}
    with respect to both the driving parameters $c_\eta$ and MPS parameters $x_k$. 
    This can be done using Euler-Lagrange equations since ${\cal S}$ is a functional depending on all $c_\eta(t)$ and $x_k(t)$
    \begin{equation}
        \begin{split}
            \partial_{c_\eta}S=0;&\quad\forall\eta\in\{1,\dots,n\},\\
            \partial_{x_k}S-\frac{\rm d}{{\rm d}t}\partial_{\dot{x}_k}S=0;&\quad\forall k\in\{1,\dots,N\},
        \end{split}
    \end{equation}
    where $N$ is the number of MPS parameters $x_k$ (again, in our example this was 2). 
    Here $\dot{x_k}=\partial_tx_k$ is the derivative with respect to time. 
    Note that $S$ does not depend on $\dot{c}_\eta$ and as a result that part yields the same equations that we had already seen in case of time-local optimization. 
    Due to this, all results for $c_\eta$ as a function of MPS parameters obtained analytically can also be applied to global trajectory optimization.
    This of course still requires finding the solution to the second set of equations.
    Of course, since we pick the initial and final points we know the initial and final values of the MPS parameters ($x_k(0)$ and $x_k(\tau)$ respectively). 
    Still, solving the resulting boundary value problem exactly is generally not feasible, even numerically.
    This is particularly true for cases where $N$ is relatively large, cases where $N$ is small can be treated numerically using finite elements or similar methods. 
    
    Alternatively, one can use the approach we introduced in this work, namely we can optimize the trajectory with respect to a much smaller number of parameters.
    One way to achieve this is by expanding the trajectory in some functional basis and then truncating that basis to a finite set of functions, as we had done. 
    Essentially, we optimize in some finite space rather than the infinite space of all functions.
    Even in this case however, when the dimensionality of the space is large enough it may be difficult to find the global minimum, nevertheless, we can find local minimums through various methods, such as gradient descent. 
    This may be sufficient to find reasonably good trajectories, depending on the choice of MPS manifold, driving operators and trajectory parametrization. 
    Importantly, one can use the integrated leakage for the cost function which is considerably computationally cheaper to calculate than full quantum evolution, even when using MPS algorithms, since no singular value decompositions are computed. 
    
    \section{Floquet scars in the Ising model} 
    \label{app:ising}
    
    In this section we provide additional details regarding the calculation for the transverse and longitudinal field Ising model (TLFIM). 
    
    \begin{figure}[b]
        \centering
        \includegraphics[width=1.0\linewidth]{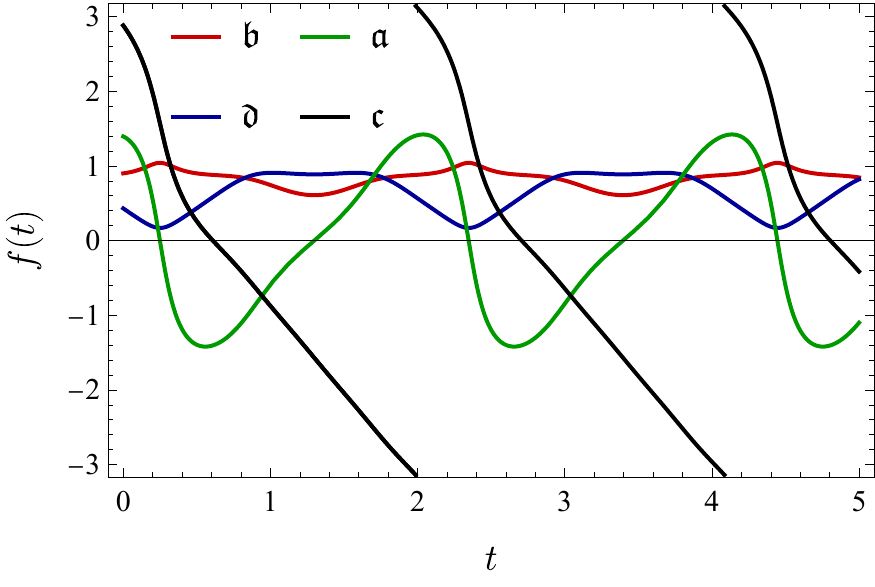}
        \caption{{
            Time dependence of the MPS parameters that define the trajectory within the MPS manifold as computed from TDVP for the TLFIM model with $J=1$, $h_x=1$ and $h_z=0.4$. 
            Note that the quick change in parameters $\frak{a}$ and $\frak{c}$ at $t\approx0.25+kt_0~;~k\in\mathbb{Z}$ corresponds to the peaks in the driving parameters observed in Fig.~\ref{fig:ising}(a) in the main text. 
        }}
        \label{sfig:isingparam}
    \end{figure}
    
    In order to obtain the periodic trajectory we use the operator $A_{\rm TLFIM}$  defined in Eq.~\eqref{eq:ising} and the MPS manifold defined in Eq.~\eqref{eq:isingmps}, choose an initial position listed in Sec.~\ref{sec:ising} and apply TDVP in order to obtain the projection of the unitary dynamics onto the MPS manifold when using time-independent parameters $J=1$, $h_x=1$ and $h_z=0.4$. 
    In this sense we follow the procedure from Ref.~\cite{Michailidis2020}, which already performed TDVP for this model. 
    Doing so we obtain the values of the 4 MPS parameters $\frak{a}$, $\frak{b}$, $\frak{c}$ and $\frak{d}$ as functions of time which we show in Fig.~\ref{sfig:isingparam}. 
    Note that the period of the trajectory in this case is $t_0\approx2.097$. 
    
    We compare the performance of the leakage minimization approach to that of trace CD and ground state CD approaches. 
    The application of CD approach requires the parent Hamiltonian for our MPS. 
    Constructing the parent Hamiltonian is relatively straightforward~\cite{Fannes1992,PerezGarcia2007}.
    We first choose the support of the local Hamiltonian density, generally it is guaranteed that a solution will exist so long as $\chi^2<d^{l_s}$, where $\chi$ is the bond dimension of the MPS, $d$ is the local Hilbert space dimension and $l_s$ is the support of the Hamiltonian density. 
    In some cases one may find solutions for smaller supports, but that is not the case in our example, so we choose $l_s=3$, which is the first integer that satisfies the inequality. 
    
    We then construct a set of vectors 
    \begin{equation}
        V=\left\{\sum_{\underline{s}}\left(\tr\left[X\prod_{i=1}^{l_s}M^{s_i}\right]|\underline{s}\rangle\right);X\in\{\mathbb{R}^{\chi\times\chi}\}\right\},
    \end{equation}
    where $\underline{s}$ is the string of physical states of the $l_s$ sites and the sum runs over all possible configurations and $X$ runs over the basis of $\chi\times\chi$ matrices. 
    We now find the complement of the linear space spanned by elements of $V$ with respect to the full Hilbert space of the $l_s$ sites, which we will denote with $W$. 
    Constructing the local Hamiltonian density is then accomplished by simply reorganizing the vectors from $W$ into a matrix $W'$ such that the individual rows correspond to the elements of the complement $W$ and computing
    \begin{equation}
        h_{\rm density}={W'}^\dagger\Lambda W'.
    \end{equation}
    Here $\Lambda$ is a diagonal square matrix of dimension equal to $|W|$ (the cardinality of the set $W$) with elements $\lambda_i>0$. 
    In this way we construct a projector which projects away from the MPS state (i.e. the state of the MPS is in the kernel of $h_{\rm density}$). 
    The parameters $\lambda$ are free parameters in the parent Hamiltonian, that can be changed while keeping a desired MPS a ground state of the Hamiltonian. 
    Such a local Hamiltonian density is guaranteed to have the MPS as a unique zero energy ground state with a spectral gap~\cite{Fannes1992,PerezGarcia2007}, yielding a parent Hamiltonian. 
    Unfortunately the resulting expressions are too bulky to be included here but we note that it is a combination of terms of the form $\frak{e}_{i-1}\otimes\frak{f}_i\otimes\frak{g}_{i+1}$, where $\frak{e}_i,\frak{g}_i\in\{\id_i,\sigma_i^z\}$ and $\frak{f}_i\in\{\id_i,\sigma^x_i,\sigma^y_i,\sigma^z_i\}$. 
    
    Using the parent Hamiltonian, we perform the calculation for both CD approaches following a similar procedure as for the PXP model. 
    The only difference is that we have a full translational invariance in the problem, corresponding to a single site unit cell. 
    At this point it is worth emphasizing again that presence of free parameters $\lambda_i$ leads to ambiguity, as the results depend on the choice of these parameters. 
    Furthermore, choosing a larger support will naturally increase the number of these free parameters which at present can be considered to be equal to zero. 
    
\begin{figure}
        \centering
        \includegraphics[width=1.0\linewidth]{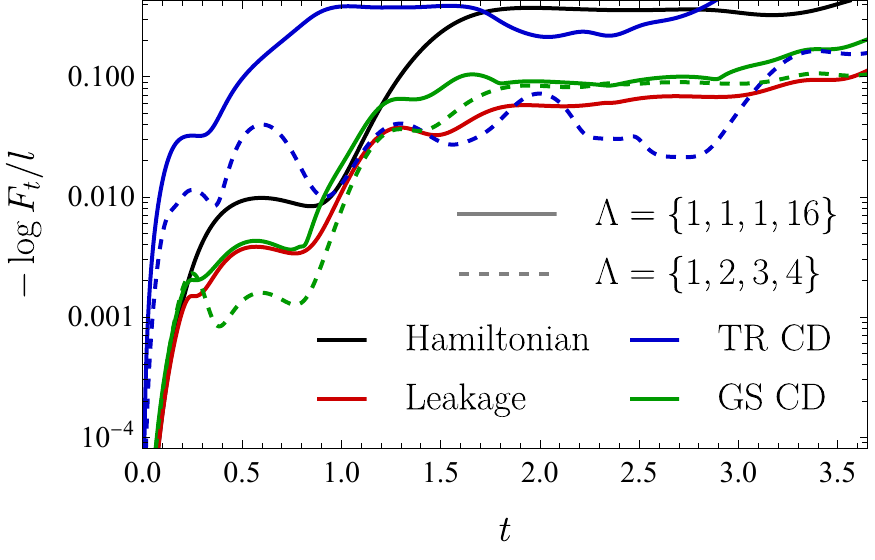}
        \caption{{
            The logarithm of the fidelity for the Hamiltonian TLFIM dynamics (black) and the three optimization approaches (colored). 
            For the CD approaches the full lines correspond to a choice of free parameters $\Lambda=\{1,1,1,16\}$ while the dashed lines correspond to $\Lambda=\{1,2,3,4\}$. 
            We observe that there is considerable difference between the two choices, and, in line with expectations, the leakage approach outperforms CD approaches at early times.
        }}
        \label{sfig:isingcomp}
    \end{figure}

    In Fig.~\ref{sfig:isingcomp} we show the performance of the different approaches for two fixed choices of free parameters $\lambda=\{1,2,3,4\}$ (dashed lines) and $\lambda=\{1,1,1,16\}$ (full lines). 
    We note here that the leakage-based approach is locally optimal, and hence it is guaranteed to perform the best at early times, as we see on the plot. 
    Later the CD approach may lead to better performance, however such instances are atypical as they rely on the fact that exact quantum dynamics returns to the MPS manifold, featuring non monotonic dependence of fidelity. 
    It is worth noting that the CD approaches are local in time and thus do not predict these spontaneous returns to the MPS trajectory. 
    Different CD behavior can be illustrated by choosing different values for the $\Lambda$ parameters. 
    Full lines in Fig.~\ref{sfig:isingcomp} show more typical behavior of CD performance, whereas the dashed lines illustrates the choice of $\Lambda$ which features such an accidental~(i.e.~not guaranteed by any symmetries or physical principles) return to the MPS trajectory. 
    
    \bibliography{newbib}

\end{document}